\documentclass[useAMS,usenatbib]{mnras}
\usepackage{float}
\usepackage{mathtools}

\usepackage[T1]{fontenc}
\usepackage{ae,aecompl}
\usepackage[caption=false]{subfig}

\usepackage{graphicx}	
\usepackage{amsmath}	
\usepackage{amssymb}	
\usepackage{multicol}        
\usepackage{bm}		
\usepackage{pdflscape}	

\usepackage[T1]{fontenc}
\usepackage{ae,aecompl}

\title[\textit{Mass-radius relation of M82 disk SSCs}]{Mass-radius relation of intermediate-age disk super star clusters of M82}

\author[Cuevas-Otahola et al.]{
B. Cuevas-Otahola,$^{1}$\thanks{E-mail: bolivia@inaoep.mx}
Y. D. Mayya,$^{1}$
I. Puerari$^{1}$
and D. Rosa-Gonz\'alez$^{1}$
\\
$^{1}$Instituto Nacional de Astrof\'isica, \'Optica y Electr\'onica, 72840 Puebla, Mexico\\
}

\date{Last updated 2020 Nov 10; in original form 2020 May 24}

\pubyear{2019}

\begin{document}
\label{firstpage}
\pagerange{\pageref{firstpage}--\pageref{lastpage}}
\maketitle

\begin{abstract}
We present a complete set of structural parameters for a sample of
99 intermediate-age Super Star Cluster (SSCs) in the disk of M82,
and carry out a survival analysis using the semi-analytical cluster
evolution code EMACSS.
The parameters are based on the profile-fitting analysis carried out
in a previous work,
with the mass-related quantities derived using a mass-to-light ratio for a
constant age of 100~Myr. The SSCs follow a power-law mass function
with an index $\alpha=1.5$,
and a log-normal size function with a typical half-light radius,
$R_{\rm h}$=4.3~pc,
which are both comparable with the values for clusters in the
Magellanic Clouds, rather than in giant spirals.
The majority of the SSCs follow a power-law mass-radius relation with
an index of $b=0.29$$\pm$0.05.
A dynamical analysis of M82 SSCs using EMACSS suggests that 23\% of the
clusters are
tidally-limited, with the rest undergoing expansion at present.
Forward evolution of these clusters suggests that the majority would
dissolve in $\sim$2~Gyr.
However, a group of four massive compact clusters, and another group
of five SSCs
at relatively large galactocentric distances, are found to survive for
a Hubble time.
The model-predicted mass, $R_{\rm h}$, $\mu_V$ and core radius of
these surviving SSCs
at 12~Gyr are comparable with the corresponding values for the sample
of Galactic globular clusters.
\end{abstract}





\section{Introduction}

Stability of auto-gravitating systems such as star clusters is controlled by
the relation between their mass (M), radius (R) and velocity dispersion ($\sigma$) \citep{Spitzer_book_gcs}.
For virialised systems, these three quantities are related by the equality
$\sigma^2 R \propto M$.
Evolution of stars in clusters and the dynamical evolution of clusters change the 
mass and $\sigma$ independently, thus forcing a change of their radius \citep{Gieles2010}.
The final evolutionary fate of the clusters is dictated by the cluster's ability to
maintain virial equilibrium throughout its lifetime.
The presence of globular clusters (GCs), the oldest objects in the Universe, suggests that
at least some clusters were able to maintain this equilibrium.
These surviving clusters are found to have inter-relation between two of these 
three quantities, which is manifested by the fundamental plane relations \citep{Djorgovski1994, McLaughlin2005}.
Elliptical galaxies, which are also auto-gravitating systems but on much larger
scale, also follow fundamental plane relations \citep{Djorgovski1987}. The most well-known relation among these
is the Faber-Jackson relation (Luminosity$\sim\sigma^4$), which implies a mass-radius
relation of the form $R\propto M^b$ with $b$=0.5, under virial equilibrium conditions \citep{FaberJackson1976}.
The distribution of the half-light radius ($R_{\rm h}$) of Galactic GCs, on the other hand, 
peaks $\sim$2.5~pc, almost without any dependence on their masses \citep{Harris1996}.
\citet{Gieles2010} and \citet{Gieles2013} attribute such a behaviour to the expansion of 
clusters during early dynamical evolution. These studies found that massive GCs ($M>10^6$~M$_\odot$)
are less prone to expansion and retain the mass-radius relation with $b$=0.5.

\citet{Terlevich2018} analysed the long-term evolutionary behaviour of Young Massive Clusters (YMCs)
in magnitude ($M_{\rm B}$) vs central velocity dispersion ($\sigma_0$) plane.
They found that the evolved YMCs show a break in the $M_{\rm B}$ vs $\sigma_0$ 
relation with the mass at the break corresponding to
$M=10^6$~M$_\odot$, with systems more massive than this value following the relation
for elliptical galaxies, and lower mass systems following the relation for the GCs.
GCs being the oldest clusters, they sample the properties of surviving clusters. 
In order to understand the general evolutionary behaviour of clusters, it is mandatory to study
also the sample of clusters that do not survive for Hubble time. 
We are likely to encounter such clusters in samples of clusters
of intermediate ages ($\sim10^8$--$10^9$~yr). 
At these ages cluster evolution is expected to be dominated by the early expansion.
The size of the expanding clusters is eventually limited by the tidal radius ($R_{\rm t}$), 
which in galaxies with flat rotation curve is given by $R_{\rm t}\propto R_{\rm g}^{2/3} M^{1/3}$,
where $R_{\rm g}$ is the galactocentric radius. Thus, tidally-limited clusters located
at similar $R_{\rm g}$ are expected to follow a mass-radius relation with $b\sim$0.33. 

Many efforts had been made to observationally obtain the 
relation between the radius containing half the mass
and the cluster mass, i.e. the
mass-radius relation
for clusters younger than GCs in nearby galaxies.
\citet{FallChandar2012} found $b$=0.5 for intermediate age ($10^8$--$10^9$~yr) 
massive clusters in the LMC, SMC and other Milky Way satellites.
On the other hand, massive clusters in NGC7252 and NGC1316 support $b$=0.3 
\citep{BastianCabrera2013,Kissler2006,BastianSaglia2006,Maraston2004}.
Even shallower relation ($b<0.3$) have been found in a variety of galaxies:
in M31 \citep{Barmby2009}, in a sample of spiral galaxies \citep{Larsen2004b}, 
in the nearby interacting galaxies NGC 5194/5195 (M51) \citep{Bastian2005,Lee2005,Hwang2010},
and in the merger galaxy NGC3256 \citep{Zepf1999}. 

M82 disk clusters offer a great opportunity to understand the evolution of 
clusters at intermediate ages. The disk has a rich population of SSCs that span 3 orders
of magnitude in mass ($\sim4\times10^3$--$4\times10^6$~M$_\odot$), distributed in
relatively small range of $R_{\rm g}$ (0.5--4.0~kpc) \citep{Mayyacat}. These clusters were
formed in a disk-wide burst following its fly-by interaction with M81 $\sim$500~Myr
ago \citep{Yun1999, Mayya2006}. Spectroscopically determined ages of SSCs show a peak $\sim$150~Myr, with 
relatively narrow age range (50--350~Myr) \citep{Konst2009}. Use of SSP models indicates that the error in the mass 
determined assuming a constant age of 100~Myr introduced due to the age spread at 
the most amounts to a factor of 2.5, which is small compared to the 1000-fold
range of masses \citep{BruzualCharlot2003}. The galaxy is relatively nearby \citep[3.63~Mpc;][]{Freedman},
which allows accurate determination of structural parameters using the HST/ACS images.
Such an analysis has already been carried out by \citet[][hereinafter Paper I]{Cuevas2020}. We use this latter
dataset to study the mass-radius relation at intermediate ages.

In \S\ref{Sec:Sample}, we briefly describe the properties of the sample of SSCs in the disk of M82 and present the obtained distributions of masses and half-light radius.  In \S\ref{Sec:massrad} we discuss the mass-radius relationship at different masses and radii, as well as the relation between surface brightness and core radius for our sample SSCs. We also perform some analytical simulations to understand the relation between these SSCs and the observed GCs. We discuss and summarise our results in \S\ref{Sec:Conclusions}.

\begin{table*}
\small\addtolength{\tabcolsep}{-3pt}
\begin{center}
\caption{Moffat-EFF model-derived parameters}
\label{tab:der_pars}
\begin{tiny}
\begin{tabular}{lrrrrrrrrrrrrrr}
\hline
ID & $\rm R_g$ & $\rm \gamma$ &  $\rm R_c$ &  $\rm R_h$ &  $\rm R_t$ &  $\rm \frac{M_{b}}{M}$ & $\rm log(\rho_0)$ & $\rm log(\Sigma_0)$ & $\rm log(M)$ & $\rm log(Ltot)$  & $log(\rho_h)$ & $log(I_h)$  &  $\rm \sigma_{p,0}$ & $\frac{\rm R_{h,0}}{\rm R_{j,0}}$\\
 & (kpc) & (pc) & (pc) & (pc) & & &  ($\rm M_\odot/ pc^{3}$) & ($\rm M_\odot/ pc^{2}$) &  ($\rm M_\odot$) & $\rm L_\odot$ & ($\rm M_\odot/pc^3$) & ($\rm L_\odot/pc^2$)  & ($\rm km/s$) &  \\
(1) & (2) & (3) & (4) & (5) & (6) & (7) & (8) & (9) & (10) & (11) & (12) & (13) & (14) & (15)\\
\hline
D1 & 0.70 & 2.73$_{-0.03}^{+0.04}$ & 0.57$_{-0.03}^{+0.05}$ & 1.67$_{-0.09}^{+0.15}$ & 31.67$_{-0.14}^{+0.29}$ & 0.94 & 4.99$_{-0.03}^{+0.04}$ & 5.06$_{-0.15}^{+0.19}$ & 5.68$_{-0.13}^{+0.28}$ & 6.57$_{-0.13}^{+0.28}$ & 4.39$_{-0.01}^{+0.06}$ & 5.32$_{-0.01}^{+0.06}$ & 13.35$_{-0.14}^{+0.29}$ & 0.04\\
D8 & 2.61 & 3.39$_{-0.02}^{+0.02}$ & 2.21$_{-0.04}^{+0.04}$ & 4.07$_{-0.02}^{+0.02}$ & 11.76$_{-0.07}^{+0.07}$ & 0.85 & 2.33$_{-0.05}^{+0.05}$ & 2.98$_{-0.07}^{+0.07}$ & 4.62$_{-0.04}^{+0.04}$ & 5.51$_{-0.04}^{+0.04}$ & 2.17$_{-0.04}^{+0.04}$ & 3.49$_{-0.04}^{+0.04}$ & 2.29$_{-0.06}^{+0.06}$ & 0.06\\
D23 & 3.75 & 3.43$_{-0.03}^{+0.01}$ & 1.01$_{-0.03}^{+0.02}$ & 1.83$_{-0.04}^{+0.02}$ & 37.77$_{-0.09}^{+0.04}$ & 0.99 & 3.71$_{-0.06}^{+0.03}$ & 4.03$_{-0.10}^{+0.05}$ & 4.98$_{-0.05}^{+0.03}$ & 5.87$_{-0.05}^{+0.03}$ & 3.57$_{-0.00}^{+0.00}$ & 4.54$_{-0.00}^{+0.00}$ & 5.78$_{-0.07}^{+0.03}$ & 0.01\\
D50 & 0.86 & 3.85$_{-0.06}^{+0.10}$ & 1.71$_{-0.07}^{+0.11}$ & 2.75$_{-0.05}^{+0.08}$ & 11.97$_{-0.15}^{+0.23}$ & 0.94 & 3.62$_{-0.08}^{+0.10}$ & 4.17$_{-0.13}^{+0.16}$ & 5.53$_{-0.05}^{+0.08}$ & 6.42$_{-0.05}^{+0.08}$ & 3.59$_{-0.01}^{+0.01}$ & 4.74$_{-0.01}^{+0.01}$ & 8.12$_{-0.11}^{+0.17}$ & 0.06\\
D51 & 2.20 & 3.99$_{-0.20}^{+0.24}$ & 3.94$_{-0.32}^{+0.39}$ & 6.13$_{-0.12}^{+0.15}$ & 5.53$_{-0.55}^{+0.67}$ & 0.45 & 1.80$_{-0.12}^{+0.13}$ & 2.71$_{-0.13}^{+0.14}$ & 4.78$_{-0.14}^{+0.17}$ & 5.67$_{-0.14}^{+0.17}$ & 1.79$_{-0.21}^{+0.26}$ & 3.29$_{-0.21}^{+0.26}$ & 1.71$_{-0.40}^{+0.49}$ & 0.09\\
D163 & 1.06 & 4.00$_{-1.70}^{+0.10}$ & 4.53$_{-1.67}^{+0.15}$ & 7.04$_{-0.71}^{+0.05}$ & 5.91$_{-2.68}^{+0.24}$ & 0.41 & 2.51$_{-0.26}^{+0.07}$ & 3.48$_{-0.27}^{+0.08}$ & 5.68$_{-0.21}^{+0.05}$ & 6.56$_{-0.21}^{+0.05}$ & 2.51$_{-0.03}^{+0.08}$ & 4.07$_{-0.03}^{+0.08}$ & 4.23$_{-1.90}^{+0.17}$ & 0.11\\
D199 & 1.30 & 4.00$_{-0.31}^{+0.45}$ & 1.20$_{-0.15}^{+0.21}$ & 1.86$_{-0.19}^{+0.27}$ & 17.38$_{-0.50}^{+0.71}$ & 0.99 & 3.84$_{-0.20}^{+0.24}$ & 4.23$_{-0.37}^{+0.45}$ & 5.27$_{-0.10}^{+0.13}$ & 6.15$_{-0.10}^{+0.13}$ & 3.84$_{-0.07}^{+0.10}$ & 4.82$_{-0.07}^{+0.10}$ & 7.73$_{-0.36}^{+0.51}$ & 0.03\\
D296 & 1.25 & 4.00$_{-1.32}^{+2.16}$ & 2.82$_{-0.95}^{+1.47}$ & 4.38$_{-0.60}^{+0.95}$ & 10.55$_{-2.07}^{+3.32}$ & 0.85 & 3.21$_{-0.29}^{+0.36}$ & 3.97$_{-0.37}^{+0.47}$ & 5.75$_{-0.21}^{+0.29}$ & 6.64$_{-0.21}^{+0.29}$ & 3.21$_{-0.09}^{+0.21}$ & 4.56$_{-0.09}^{+0.21}$ & 7.07$_{-1.47}^{+2.35}$ & 0.05\\
D303 & 3.20 & 3.89$_{-0.10}^{+0.09}$ & 2.05$_{-0.12}^{+0.10}$ & 3.27$_{-0.08}^{+0.07}$ & 8.79$_{-0.25}^{+0.21}$ & 0.87 & 2.09$_{-0.10}^{+0.09}$ & 2.72$_{-0.15}^{+0.14}$ & 4.24$_{-0.08}^{+0.06}$ & 5.12$_{-0.08}^{+0.06}$ & 2.07$_{-0.04}^{+0.02}$ & 3.30$_{-0.04}^{+0.02}$ & 1.59$_{-0.18}^{+0.16}$ & 0.05\\
\hline
\end{tabular}
\end{tiny}
\hfill\parbox[t]{\textwidth}{
Description of the columns:
(1) Cluster ID.  
(2) Galacto-centric radius in kiloparsecs.  
(3) Moffat-EFF power-law index (from Paper~I except that $\gamma>$4 are set to 4).
(4) Core radius in parsecs (calculated using Eq.~\ref{eqn:rc_rd}). 
(5) Half-light radius in parsecs (calculated using Eq.~\ref{eqn:rh_rd_moffat}). 
(6) Tidal radius in parsecs (see Appendix~\ref{App:vel_disp}). 
(7) Fraction of the total mass of Moffat-EFF profile within the tidal radius (calculated using Eq.~\ref{eqn:Mbound}).
(8-13) Logarithm of central mass volume density, central mass surface density,
total mass, total luminosity, half-light mass volume density, and average surface brightness within $R_{\rm h}$, respectively.
The mass-related quantities (columns 8, 9, 10, 12 and 14) are obtained from the corresponding luminosity-related 
quantities assuming a mass-to-light ratio for an SSP of 100~Myr and using a Kroupa IMF \citep{KroupaIMF}.
The effects of a different age choice in the values in columns 8, 9, 10, and 12 are given 
by adding the term $0.57\log(t/{100\, Myr})$ and in column 14 by multiplying by $\sqrt{(t/{100\, Myr})^{0.57}}$.
(14) Projected central velocity dispersion (see Appendix~\ref{App:vel_disp}). 
(15) Initial half-mass to Jacobi radius ratio.
This table is shown in its entirety in the electronic edition.  
A portion is shown here for guidance, which is constituted by the most interesting set of SSCs (see \S\ref{Sec:long}).
}
\end{center}
\end{table*}

\begin{figure}
\begin{center}
\includegraphics[width= \columnwidth]{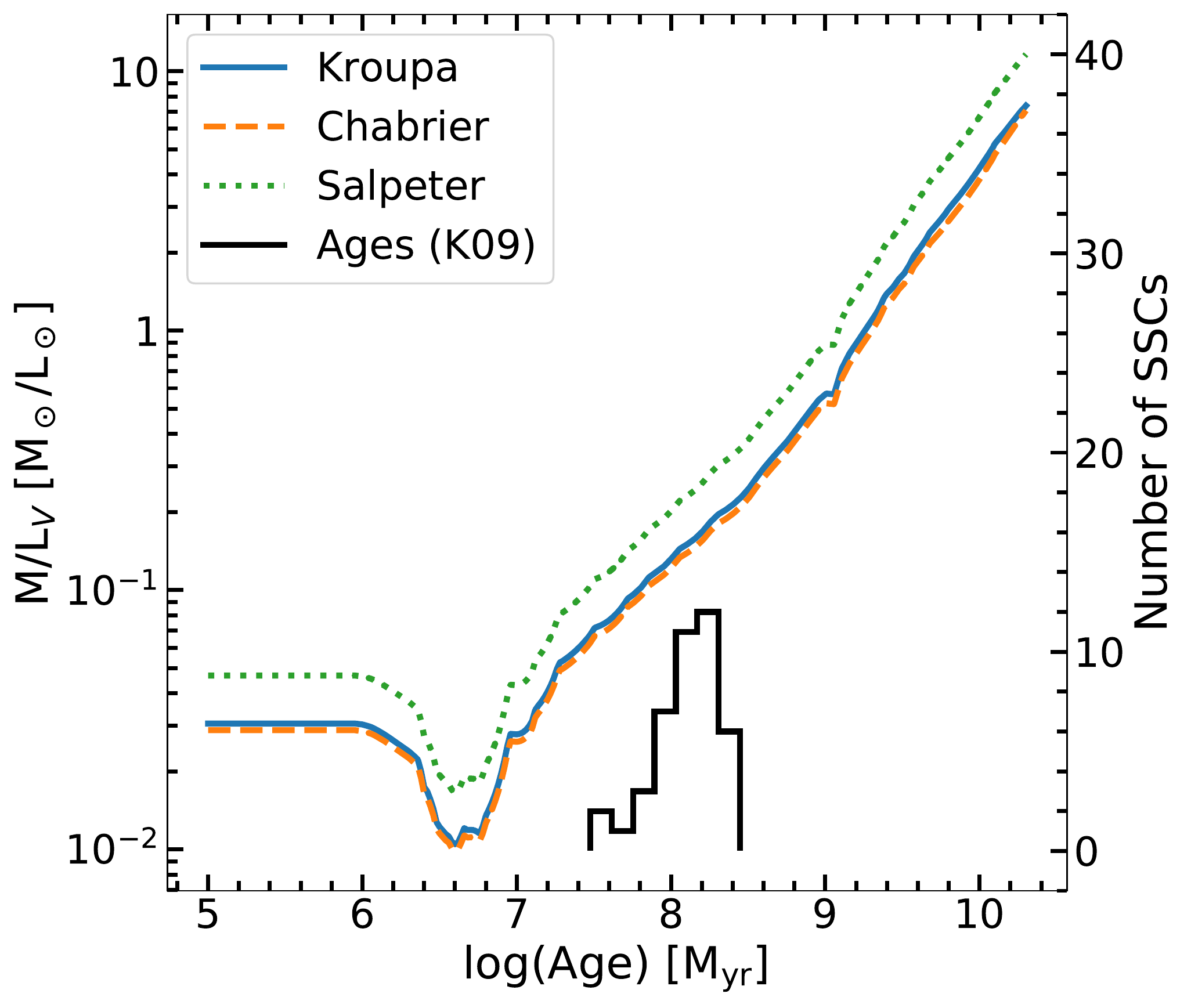}
\caption{Mass-to-Light ratio in the V-band versus log(Age) for 
\citet{BruzualCharlot2003} (BC03) Simple Stellar Populations (SSPs)  with solar 
metallicity for Kroupa \citep{KroupaIMF} (solid blue line), Chabrier 
\citep{ChabrierIMF} (orange dashed line) and Salpeter \citep{SalpeterIMF} (green 
dotted line) Initial Mass Functions (IMFs).  The three lines are parallel, with 
the Salpeter IMF having the higher Mass-to-Light ratios.  The black histogram 
(numbers on the right axis) represents the age distribution for M82 disk SSCs 
reported by \citet{Konst2009}. }
\label{fig:bc_03_edad}
\end{center}
\end{figure}

\section{Sample and the derived parameters}\label{Sec:Sample}

In this work, we analyse the mass-radius relation for the M82 disk SSCs sample studied in Paper I. 
The sample consists of 99 SSCs from the M82 disk SSCs sample of 393 clusters 
from \citet{Mayyacat}, selected on the archival images of the HST Legacy 
Survey \citep{Mutchler2007}.  In Paper I, we have demonstrated that the 
sub-sample of 99 SSCs represents the bright end of the total sample.  
The structural parameters were derived for the best-fitting Moffat-EFF 
\citep{Elson}, King \citep{King_dyn} and Wilson \citep{Wilson_dyn} models.
We found that the Moffat-EFF model is the best-fit for 95\% of the cases,
and hence in this study we use the parameters of this model. The Moffat-EFF model
had been used in the characterisation of clusters of intermediate age in 
other galaxies, e.g. the LMC, SMC \citep{MackeyGilmore2003a,MackeyGilmore2003b}, 
Antennae \citep{Whitmore&Schweizer1995}, M83 and  NGC1313/628 \citep{Ryon,Ryon2017}.    

In Paper~I, we presented the results for the core radius $R_{\rm c}$ and 
the power-law index $\gamma$ for the disk sample. In this paper, we 
use the results of the profile fitting to calculate 
the full set of structural parameters, that includes 
half-light radius $R_{\rm h}$, central mass density ($\rm log\rho_0$),
central surface density ($\rm log\Sigma_0$), central velocity dispersion, 
luminosity and mass. We include also an important quantity 
related to the initial conditions of the cluster with respect to the galaxy
tidal field, i.e., the initial half-mass to Jacobi radius ratio $\frac{R_{\rm h,0}}{R_{\rm j,0}}$ 
(See \S\ref{subsec:emacss}). 
Some of these parameters depend on the dynamical evolution of the cluster,
and hence this dataset is useful to test the predictions of models of 
dynamical evolution at intermediate ages.
In Table \ref{tab:der_pars}, we summarise all the derived parameters in 
the V-band for the Moffat-EFF model for all the clusters.  We note that 
not all the tabulated parameters are independent of each other. Nevertheless, 
considering their usefulness we list them.
 
\subsection{Age and extinction of the clusters}\label{Sec:ages}

Among the set of Moffat-EFF parameters, $R_{\rm c}$, $R_{\rm h}$ and $\gamma$ are directly
derived from the best-fit model, whereas the other parameters are derived 
assuming an age and extinction. In the first place, most of the disk clusters 
in M82 are believed to have formed in a disk-wide burst around 300~Myr ago 
following the last fly-by interaction with its neighbour M81 \citep{Yun1999,Mayya2006}.
Available spectroscopic ages for a sample of 42 disk SSCs \citep{Konst2009} 
are plotted in Fig.~\ref{fig:bc_03_edad} (the inset). 
The majority has ages between 100--300~Myr, with the median $\sim$150~Myr. 
The relatively small dispersion in ages supports the burst-origin scenario for the formation of disk SSCs. 
In this figure, we also show the mass-to-light ratio ($\Gamma$) variation as a function of age
using \cite{BruzualCharlot2003} Simple Stellar Population models (SSPs).
$\Gamma$ lies between 0.08 to 0.2 in the V-band for the range of ages of M82 SSCs for Kroupa \citep{KroupaIMF} 
and Chabrier \citep{ChabrierIMF} 
Initial Mass Functions (IMFs). The values are around a factor of 2.5 higher for Salpeter IMF \citep{SalpeterIMF}.
Given the burst-origin for the disk SSCs, 
all the disk SSCs, including those for which we do not have spectroscopic ages, are
expected to have ages between 100--300~Myr.
In order to determine the effects of the age in the derived parameters, we have 
fitted a power-law function of the form $(t/(100\, \rm Myr))^\alpha$ to the 
mass-to-light ratio values, with the age $t$ varying between 50--500~Myr, which 
gives $\alpha=0.57$. 
If some SSC is as young as 50~Myr, its mass would be 67\% of
the mass reported here, and if the SSC is as old as 500~Myr, its mass would
be 2.5 times higher.
Extinction is the second parameter that affects the values of some of the derived parameters.  \citet{Mayyacat} tabulated $A_V$ based on B-V and B-I colour excesses over that expected for an SSP of 100~Myr age, and using \citet{Cardelli1989} extinction curve.  We used their values here.

\subsection{Model-derived parameters}
\subsubsection{Total luminosities and masses}

From Paper I, we recall the form of the Moffat-EFF intensity profile $I(R)$: 
\begin{equation}
I(R)=\frac{(\gamma-2) L_{\rm tot}}{2 \pi r_{\rm d}^2} \bigg{[}1+\bigg{(}\frac{R}{r_{\rm d}}\bigg{)}^2\bigg{]}^{-\gamma/2}, 
\label{eq:moffat_proj}
\end{equation}
where $R$ is the semi-major axis of the observed profile, $L_{\rm tot}$
is the total asymptotic luminosity of the profile, $\gamma$ is the power-law
index, $r_{\rm d}$ is the characteristic radius which is related to the core radius, $R_{\rm c}$ by:
\begin{equation}
r_{\rm d}=\frac{R_{\rm c}}{(2^{2/\gamma}-1)^{1/2}}.
\label{eqn:rc_rd}
\end{equation}
The best-fitting model using equation~\ref{eq:moffat_proj} directly gives 
$L_{\rm tot}$, which immediately allows the 
calculation of total mass, assuming a mass-to-light ratio ($\Gamma$) value from SSP 
models for a fixed age of 100 Myr. We used a $\Gamma$ in the visual band of 0.13
corresponding to a Kroupa SSP model of \cite{BruzualCharlot2003}.

\begin{figure}
\begin{center}
\includegraphics[width= \columnwidth]{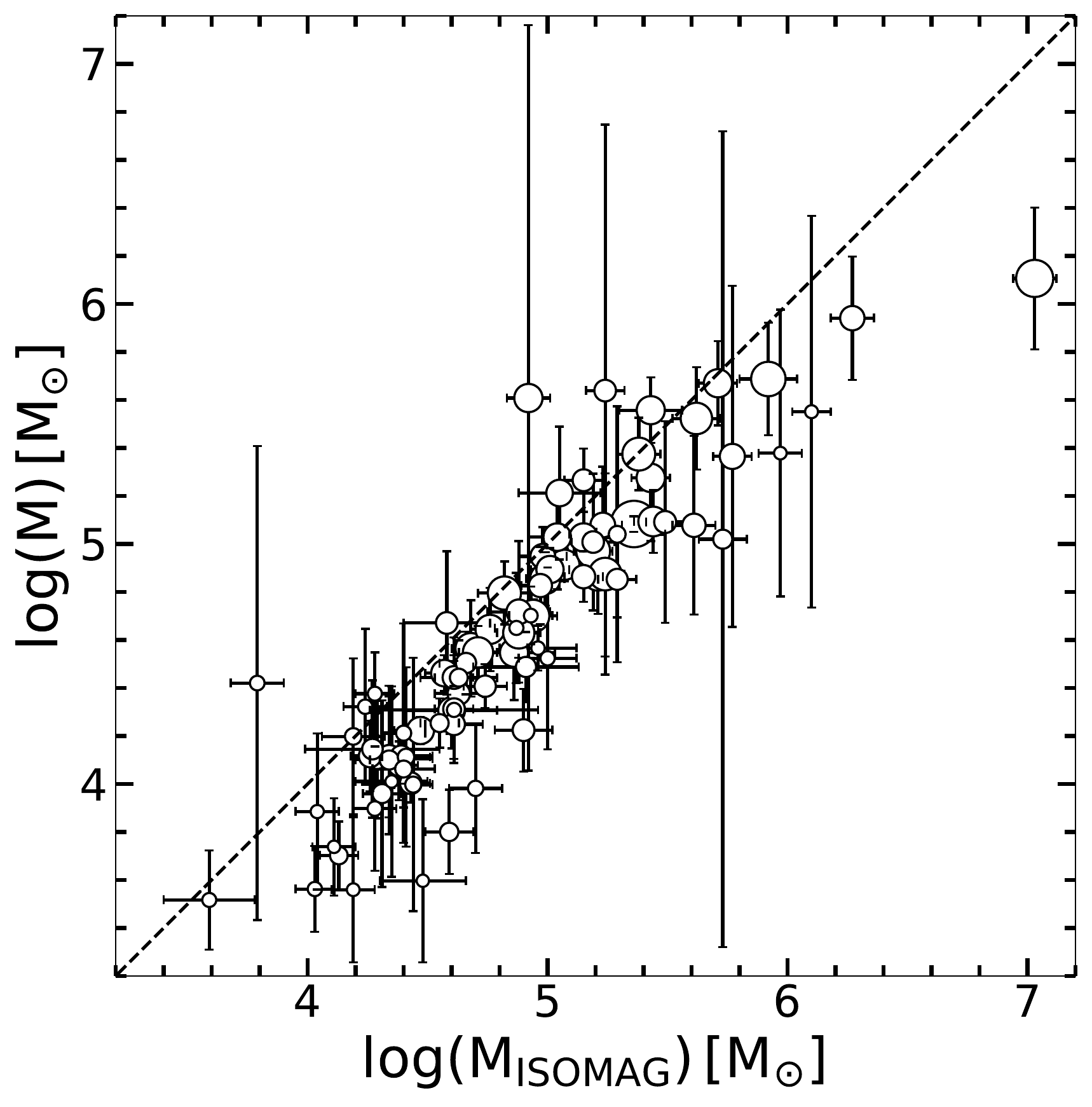}
\caption{Comparison of masses obtained from Sextractor isophote magnitude 
($\rm M_{\rm ISOMAG}$) from \citet{Mayyacat} with masses ($\rm M$) 
obtained from the best-fitting Moffat-EFF model luminosities using 
Eq.~\ref{eq:moffat_proj}. Both masses were calculated assuming an SSP with an 
age value of 100~Myr from \citet{Girardi2002}. The sizes are coded by the area 
of the aperture used in $\rm M_{\rm ISOMAG}$. The diagonal line indicates the 
line of unit slope. Isophote magnitudes systematically give $\sim$10--30\% 
higher masses, which is due to underestimation of local background in Sextractor.
}
\label{fig:com_mass_mayya}
\end{center}
\end{figure}

\cite{Mayyacat} derived photometric masses ($\rm M_{\rm ISOMAG}$) for all 
the disk clusters assuming a constant age of 100~Myr and using isophote magnitudes  
obtained by SExtractor.  
The model-derived mass $M$ is expected to be more accurate than $\rm M_{\rm ISOMAG}$, 
due to a more careful treatment of the local background for subtraction.
In Fig.  \ref{fig:com_mass_mayya}, we compare these two masses. For the sake of comparison,  
we used the \citet{Girardi2002} SSPs to be consistent with the mass calculated in \citet{Mayyacat}.
The errors on $\rm M_{\rm ISOMAG}$ (x-axis) take into account the errors in the determination of magnitude and
the extinction, whereas the errors on model-derived masses (y-axis) are obtained by propagating the errors of 
the best-fit parameters. As expected, there is good agreement between the two mass determinations. 
However, $\rm M_{\rm ISOMAG}$ masses are systematically larger by $\sim0.2$~dex. 
This is due to underestimation of background in the SExtractor-determined ISOMAG.
The masses derived here are current masses (not initial masses) and take into account mass-loss from stars during the evolution of clusters.

\begin{figure*}
\begin{center}
\includegraphics[width= 1.8\columnwidth]{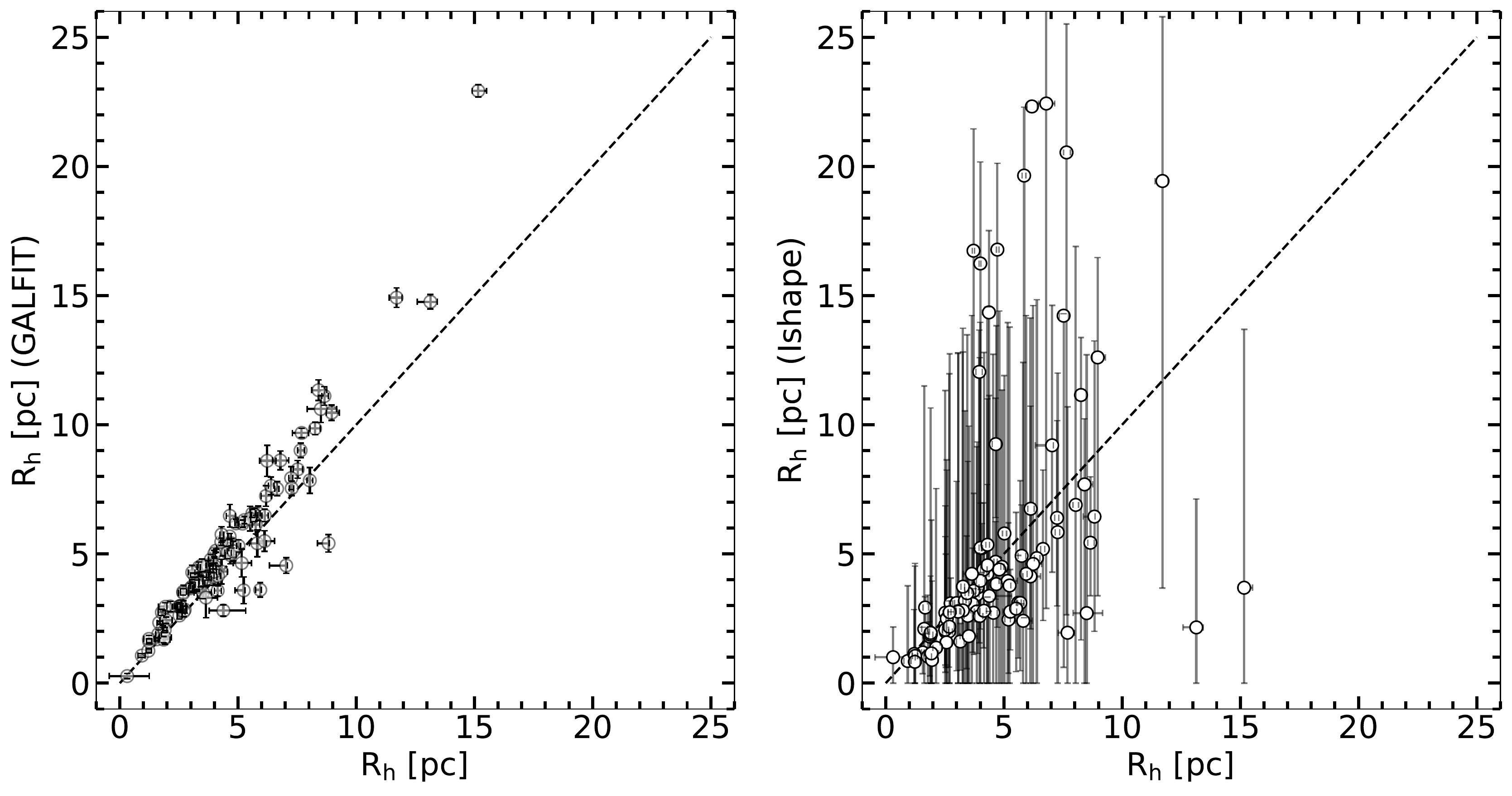}
\caption{
Comparison of $R_{\rm h}$ derived in this work using our code with those from 
{\sc galfit} (left panel) and {\sc Ishape} (right panel) obtained in Paper I. 
Error bars on the derived $R_{\rm h}$ are shown for all measurements. 
Our $R_{\rm h}$ values as well as their errors compare well with those
obtained with {\sc galfit}. Values are in general agreement with those
ontained with {\sc Ishape}, with the latter systematically having a much
larger error bars as compared to ours. 
}  
\label{fig:Rh_comparison}
\end{center}
\end{figure*}

\subsubsection{Model-derived $R_{\rm h}$, $\rho_0$, $\sigma_p$, $R_{\rm t}$}\label{subsec:pars}

The half-mass radius, the radius that contains half of the total mass is 
traditionally used as a proxy to the cluster size in most theoretical works.
For clusters that do not have strong colour gradients, it is identical to
the observationally measurable quantity, the half-light radius, $R_{\rm h}$. 
For the Moffat-EFF profile, the $R_{\rm h}$ is analytically related to the fitted
structural parameters $r_{\rm d}$ and $\gamma$ by 
\begin{equation}\label{eqn:rh_rd_moffat}
R_{\rm h} = r_{\rm d}(0.5^{1/(1-\gamma/2)}-1)^{1/2}.
\end{equation}
The errors on $r_{\rm d}$ and $\gamma$ are propagated to obtain the errors on the derived $R_{\rm h}$.
In Figure~\ref{fig:Rh_comparison}, we compare the $R_{\rm h}$ calculated using our code
with the corresponding values obtained by us using two other popularly used codes,
{\sc galfit} \citep{Penggalfit} and {\sc Ishape} \citep{Larsenishape}. For this purpose, we used the fits that we carried out
using these codes in Paper~I. As in Paper~I for $r_{\rm d}$, the agreement between $R_{\rm h}$
measurements is excellent with {\sc galfit}, whereas there is a larger dispersion of the values obtained with {\sc Ishape}.

Another parameter of interest is the central mass density $\rho_0$, 
which is related to the central luminosity density $j_0$ by $\rho_0=j_0\Gamma$. 
The $j_0$ is obtained from the best-fitting $I_0$ following the prescription
of \citet{Elson}.

Moffat-EFF profile is an empirical profile and hence does not have the underlying equations
that describe the radial density structure and stability of a cluster. 
However, the velocity dispersion profile of a spherical cluster that has a Moffat-EFF profile
can be derived under the simplifying assumptions of hydrostatic equilibrium and isotropic 
velocity distribution. 
\citet{Elson} found the solution for the velocity dispersion profile for such a 
cluster under the presence of the tidal force of the host galaxy. We used their equation 16 
to obtain velocity dispersion profile $\sigma$(r) for each one of the M82 clusters. 
These $\sigma$(r) profiles are then projected onto the plane of the sky, following the method described in 
Appendix~\ref{App:vel_disp}, where we also show the resulting $\sigma$(r) profiles 
for one illustrative cluster at various values of galacto-centric distances. 
It can be seen that the $\sigma$(r) in the presence of a tidal field drops abruptly to
zero at a finite radius, in spite of the observed intensity profiles not showing any
truncation. The radius where $\sigma$(r) reaches zero is the tidal radius $R_{\rm t}$, and hence
the contribution to the intensity outside this radius comes from unbound stars. 
The projected central velocity dispersion $\sigma_{\rm p,0}$, which is a direct
observational quantity, and $R_{\rm t}$, are tabulated
in columns 14 and 5 of Table~\ref{tab:der_pars}, respectively.
The most massive of our clusters, D1 (known in the literature as M82-F) has been 
the target of $\sigma_{\rm p,0}$ measurements.
The value calculated for this cluster from our fitting (13.35$_{-0.14}^{+0.29}$ km/s) 
agrees well with the observed values of 13.4$\pm$0.7 km/s obtained by \citet{SmithGallagher2001}
and 12.4$\pm$0.3 km/s obtained by \citet{McCrady2007}.

The presence of unbound stars just outside the tidal radius is expected at intermediate
ages because it takes 5--10 orbital periods around the parent galaxy for the unbound
stars to be stripped away \citep{Elson}. For the circular velocity of M82, orbital period varies from 
$\sim$30~Myr at 0.5~kpc to 250~Myr at 4~kpc, which implies a tidal stripping timescale of
150--300~Myr at the inner-most radius of the disk. At other galactocentric distances 
it would take $\geq$1~Gyr for tidal stripping. This explains the prevalescense of Moffat-EFF profiles
for M82 disk SSCs which are $\sim$100~Myr old.
From the obtained $R_{\rm t}$ values it is possible to compute the bound mass of the 
clusters $\rm M_{bound}$, by integrating the volume density profile $\rho(r)$ in the limits 0 and $R_{\rm t}$.

\begin{figure}
\begin{center}
\subfloat{\includegraphics[width= \columnwidth]{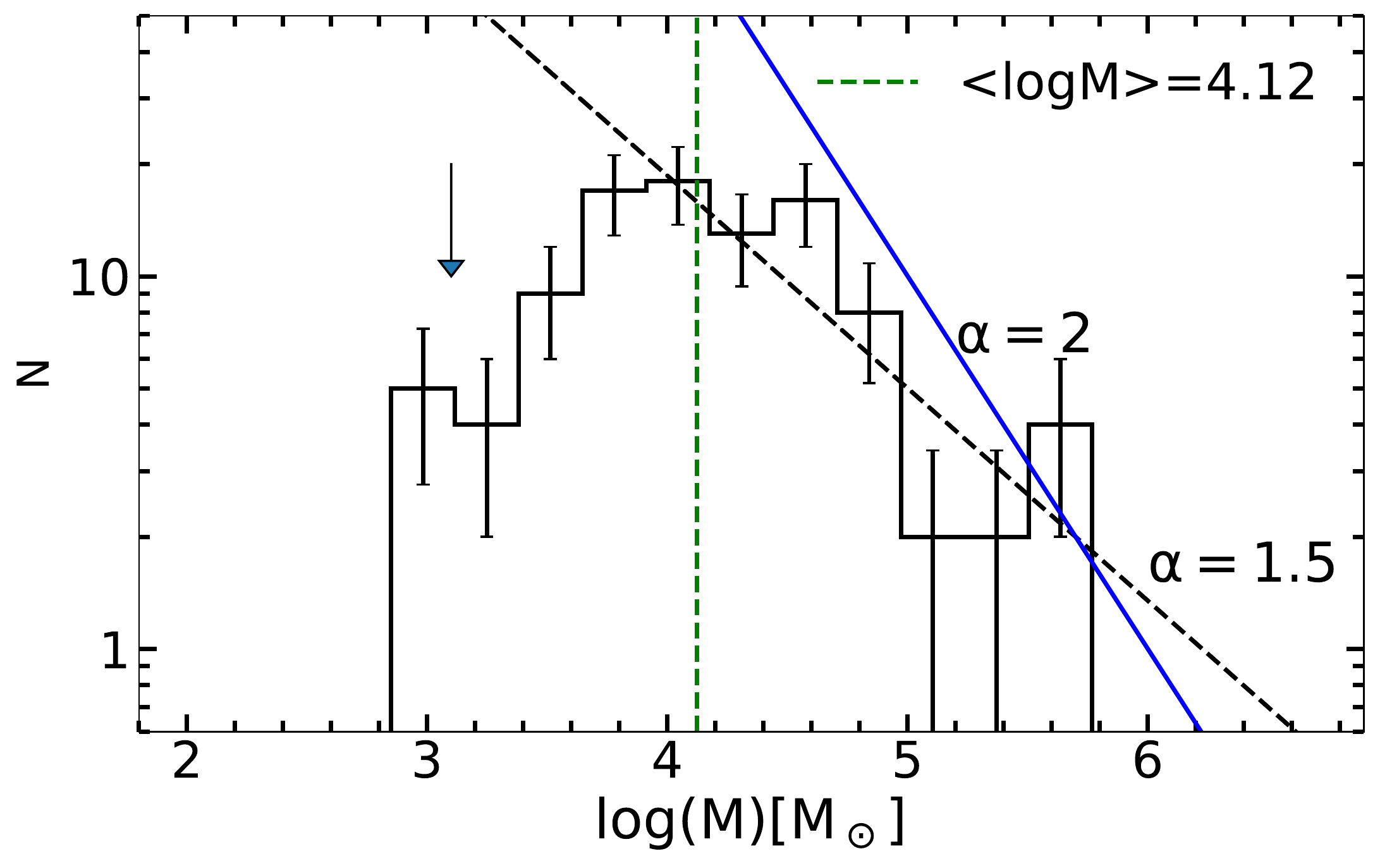}}\\
\subfloat{\includegraphics[width= \columnwidth]{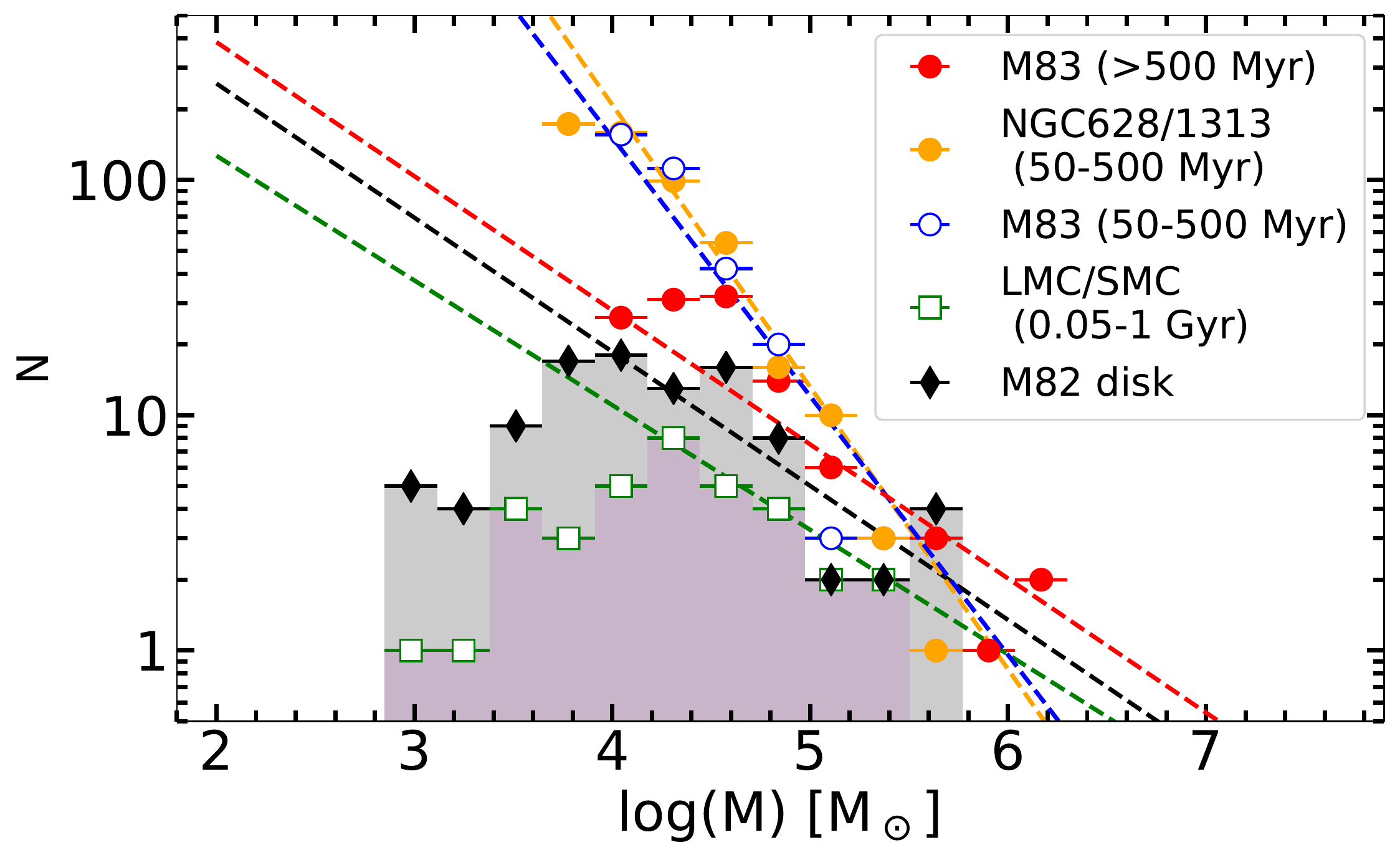}}\\
\caption{Upper panel: distribution of the total masses 
for the best-fit Moffat-EFF profile for 99 M82 SSCs (histogram). 
The error bars are computed using Poisson statistics. The green vertical 
line shows the median of the mass distribution. 
A fit to the high-mass end of the mass function is shown by the
black dashed line (power-law index of $\alpha=1.5$).
Young Clusters are characterized by $\alpha=2.0$, 
which is shown by the blue solid line.
The completeness limit is indicated with a vertical arrow.  
Bottom panel: comparison of the mass distribution for M82 disk
SSCs which are in the age range 50--300~Myr (light green histogram,
along with its power-law $\alpha=1.5$ fit in dashed black line) 
with clusters of similar age in other galaxies, which are identified 
in the inset box. Observed slope for M82 SSCs agree well with
that for the similar-age LMC/SMC clusters (purple histogram with 
the best-fit power-law shown in dashed green line), and old age
M83 clusters. On the other hand, intermediate-age clusters in 
M83 and NGC628/1313 have the same slope as for young clusters.
The horizontal bars correspond to the fixed logarithmic width used for binning.
The best-fit power-law function is shown by dashed lines following the same colour code as the binned data.
}
\label{fig:dist_sigma_pow_law}
\end{center}
\end{figure}

\subsection{Cluster mass function}\label{Sec:CMF}

Mass distribution of star clusters in galaxies is found to follow a power-law 
function. The power-law slope for young clusters is well-established and is 
close to 2 \citep{deGrijs2003} over a range of 3 orders of magnitude in mass
($\sim10^3$--$10^6$~M$_\odot$). 
Observational characterisations of the mass function for evolving populations 
(age $\geq$10~Myr) in several nearby galaxies, such as M51 \citep{Bik2003}, 
Antennae \citep{ZhangFall1999}, the starburst galaxies NGC 3310 and 
NGC 6745 \citep{deGrijsAnders2003}, and LMC \citep{deGrijs2008} do not provide 
any compelling evidence for a change, neither in form, nor in the slope of the power-law function. 
However, young SSCs in the nuclear region 
of M82 have shown a tendency for a slightly flatter slope
with \citet{McCrady2007} obtaining a slope of 1.9 and \citet{Mayyacat} obtaining a 
value of 1.8.
The distribution at the high-mass end of the CMF falls sharply, which is found
to be a common characteristic \citep{Larsen2009} and is often fitted with
a Schechter function \citep{Schechter1976}. 
Clusters loose mass during their evolution, both due to stellar evolution, and
dynamical processes \citep{Gieles_Alexander2017}. 
The former process is not expected to alter
the slope of the function, as long as the IMF of stars is independent of the mass of
the clusters, whereas the impact of the latter process is mass-dependent and hence
will influence the CMF of evolving clusters \citep{Gieles2009}. 

Availability of a large population of SSCs, all formed over a short
interval of time around $\sim$100~Myr ago in M82 disk, allows to study the CMF 
of almost a coeval population.
Such a study was carried out in \citet{Mayyacat} using the
photometric masses for the SSCs studied here, finding an $\alpha$=1.5, for
SSCs of mass above $2\times10^4~\rm M_\odot$. 
We reanalyse the CMF using the masses 
obtained from Eq. \ref{eq:moffat_proj}.
In Fig. \ref{fig:dist_sigma_pow_law}, we show the distribution of model-derived masses.  
The distribution is a power-law function with $\alpha$=1.5 for $M>10^4~\rm M_\odot$, 
reproducing the results of \citet{Mayyacat} for the subsample of clusters analysed here. The distribution remains nearly
flat between (1--4)$\times10^4~\rm M_\odot$, dropping steeply for lower masses, below the completeness limit.  

In the bottom panel of Fig. \ref{fig:dist_sigma_pow_law}, we compare the distribution of masses 
of the M82 disk SSCs with those in other galaxies where a fit of 
basic structural parameters had been carried out: LMC/SMC 
\citep{MackeyGilmore2003a,MackeyGilmore2003b}, M83, NGC1313 and NGC628 \citep{Ryon,Ryon2017}.  
The distribution and the slope of the high-mass end of the mass function for LMC/SMC intermediate-age clusters compares very well with those of M82.  On the other hand, intermediate-age SSCs in M83 and NGC628/1313 have the same slope as for young clusters ($\alpha=2$).   The slope of the high-mass end of the distribution of old clusters (>500 Myr) in M83 is similar to that in M82.   These tendencies are remarkably similar to the tendencies we have found in the distributions of $R_{\rm c}$ in Paper I.

The M82 data for disk SSCs are clearly not compatible with $\alpha$=2.0 found 
for young clusters in other galaxies. The mass function is
also flatter than that for its young clusters in the nuclear region, 
which as discussed above,
is marginally flatter than that for cluster populations in other star-forming galaxies.
Thus, even if M82 disk SSCs were formed with a flatter IMF as in the case for 
clusters in its nuclear region ($\alpha$=1.8),
the CMF has evolved, implying that the cluster evolution in the disk of M82 is mass-dependent. 
The most important dynamical process that is at work at the ages of M82 SSCs
is the tidal effect on the clusters from the gravitational potential of the
host galaxy. 
Our analysis of the structural parameters allows us to investigate this
issue, which we will carry out in \S\ref{Sec:massrad}

\subsection{$\rm R_{\rm h}$ distribution}\label{sec:Rh}

\begin{figure}
\begin{center}
\subfloat{\includegraphics[width=\columnwidth]{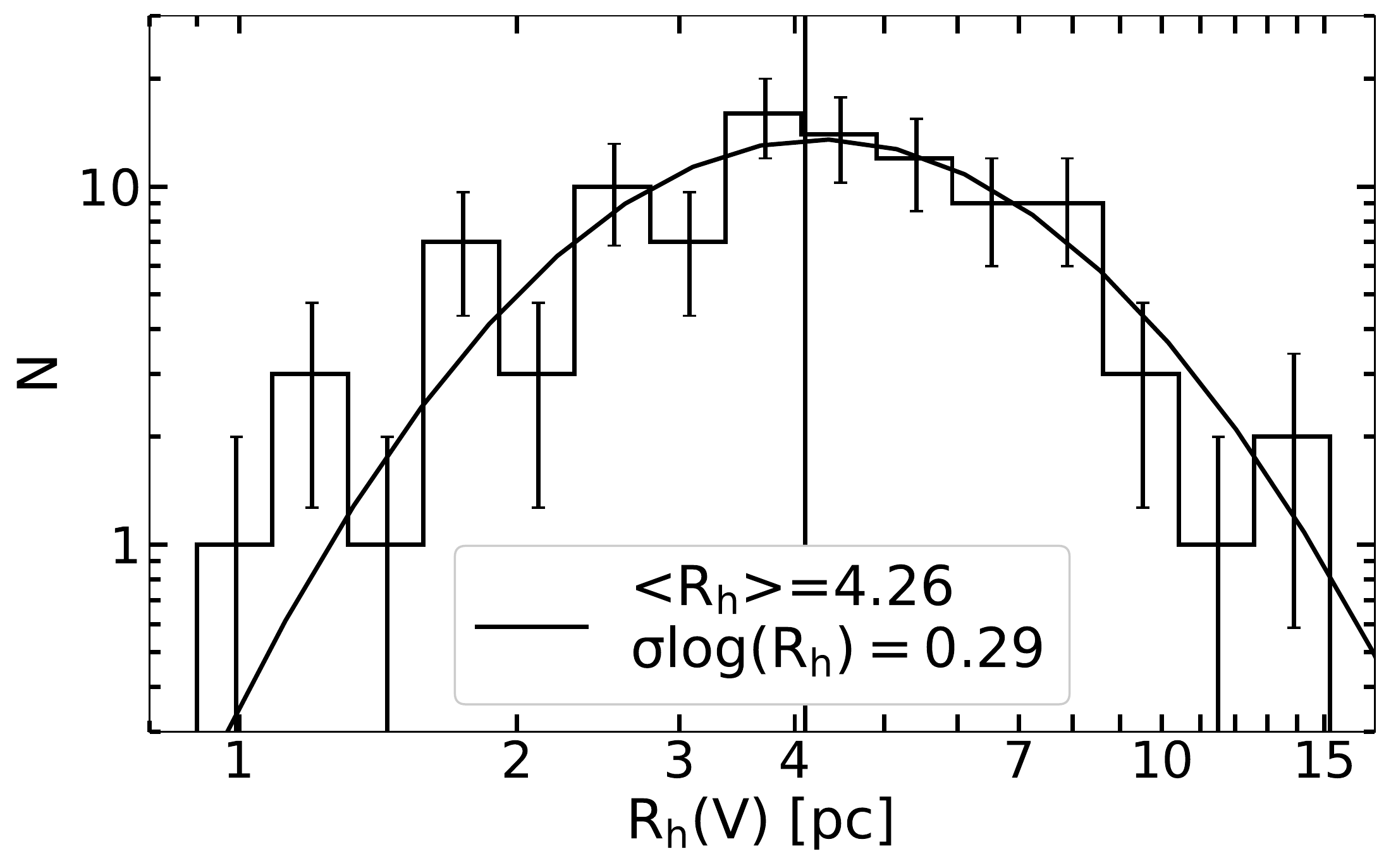}}\\
\subfloat{\includegraphics[width=\columnwidth]{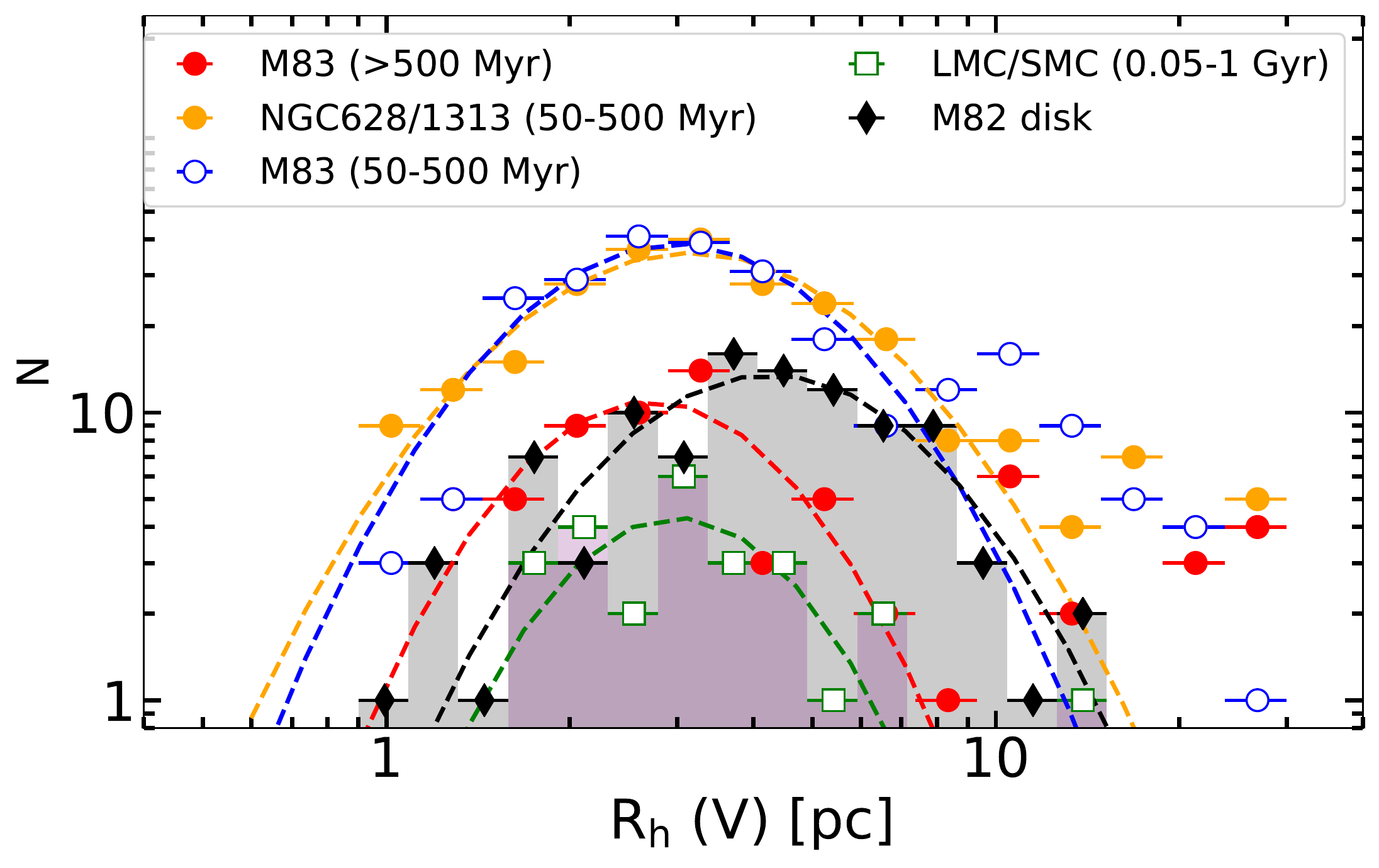}}\\
\caption{Same as Figure~\ref{fig:dist_sigma_pow_law}, but 
the distribution of $R_{\rm h}$. 
Upper panel: The $R_{\rm h}$ distribution follows log-normal function
whose center value and $\sigma$ are given in the inset box.
Bottom panel: log-normal function is a good fit in all galaxies.
However, the center of the distribution in M82 is closest to that in 
LMC/SMC, which is slightly higher than that for the rest of the 
samples.
}
\label{fig:reff_comp_moffat}
\end{center}
\end{figure}

Clusters in their early phase are expected 
to undergo adiabatic expansion leading to an increase in their radius.
This is principally driven by the loss of residual gas from the cluster volume
\citep{GoodwinBastian2006}. 
Evidence for such an expansion has been observationally found by \citet{BastianGieles2008} for
extragalactic star clusters of age $\lesssim$100~Myr.
Clusters in the LMC/SMC are also found to be in expansion \citep{MackeyGilmore2003a,MackeyGilmore2003b}.
In the presence of a tidal field, sizes of the expanding clusters are limited 
by the tidal radius \citep{King_emp}. 
At late times, contraction of the core due to core-collapse may reduce the cluster 
sizes before they reach the tidal radius.
\citet{Gieles2013} found that only a third of the Milky Way GCs have reached 
their tidal radii, with the rest still expanding. 
We here study the distribution of half-light radius for our sample of intermediate-age
SSCs in the disk of M82 and compare them with distributions of different-age clusters
in other galaxies.

In Fig. \ref{fig:reff_comp_moffat}, we show the distribution of $R_{\rm h}$ in 
the V band in logarithmic bins for the M82 disk SSCs. The distribution is well-fitted 
with a log-normal function with $\sigma \rm log(\frac{R_{\rm h}}{pc})=$~0.3~dex, 
centered at 
$R_{\rm h=}$4.26$\pm 0.26$~pc. This value is slightly higher 
than the values of $R_{\rm h}$ for young SSCs \citep{Portegiesrev}. In 
particular, for the young SSC R136 in the LMC, \citet{MackeyGilmore2003a} 
reported an $R_{\rm h}$ value as small as 1.3~pc which is a factor of 3 smaller 
compared to our mean values in M82. Our values are comparable to those of Galactic GCs. 

We compare the M82 $R_{\rm h}$ distribution with that for intermediate-age SSCs in other galaxies using the same sample described in \S\ref {Sec:CMF}.  LMC/SMC clusters have a mean value similar to that in M82, although the distribution in LMC/SMC does not show the tail towards high $R_{\rm h}$ values.  On the other hand, the $R_{\rm h}$ distribution is slightly shifted toward lower values for intermediate-age and old clusters in NGC1313/628 and M83.  These differences in the mean values of the distribution are similar to the tendencies of the $R_{\rm c}$ distribution presented in Paper I.  These differences could be related to the morphological type of the host galaxies --- giant galaxies such as M83 are expected to have intense tidal fields, which play an important role in limiting the sizes of clusters.

\begin{figure}
\begin{center}
\subfloat{\includegraphics[width= \columnwidth]{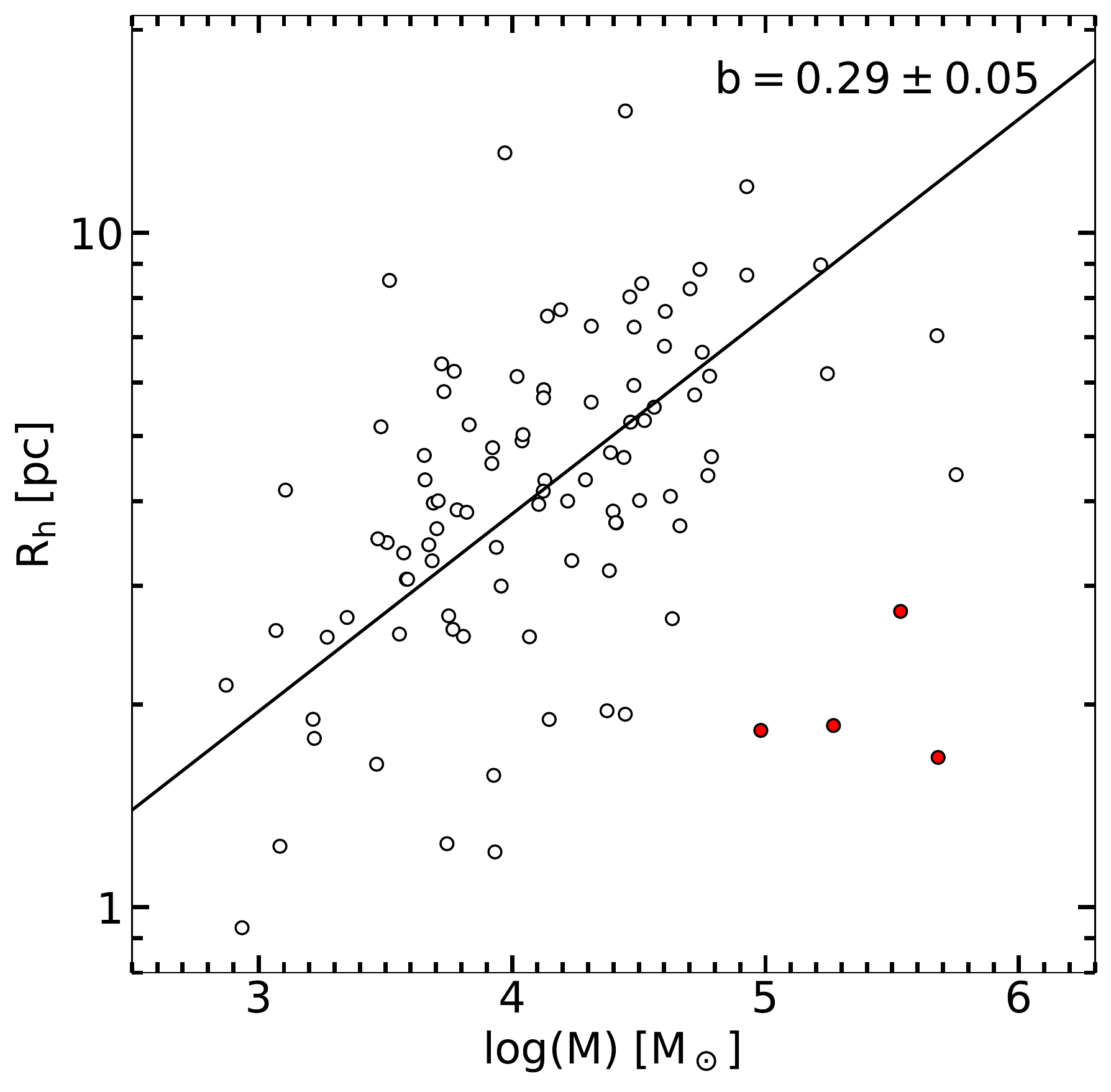}}
\caption{Radius ($\rm R_{\rm h}$) vs mass diagram for the sample of M82 disk SSCs.
The black solid line (slope $b=0.29$) shows a least-square fit of the 
mass-radius relation, excluding a group of four massive-compact SSCs 
(shown by red filled circles). 
}  
\label{fig:mass_vs_reff}
\end{center}
\end{figure}

\section{Mass-radius relation and the fundamental plane}\label{Sec:massrad}

\citet{Mayyacat} found a trend of more massive SSCs in M82 being slightly larger
than the less massive ones. In their study, they had used the Sextractor-derived
FWHM as a proxy for sizes. We here re-revisit the mass-radius relation
using the $R_{\rm h}$ obtained in this work for the sub-sample of M82 disk SSCs.

\subsection{Mass-radius relation}\label{Sec:massradius}

In  Fig.~\ref{fig:mass_vs_reff}, we plot the $R_{\rm h}$
 versus mass for the M82 disk sample analysed here. The majority 
of the SSCs (open circles) follows a trend of radius increasing with the cluster
mass over two orders of magnitude in mass. Even after taking into 
account the dispersion over this trend, we recognise a group of 
4 SSCs (shown by red filled circles), which are among the most massive 
clusters in our sample, that seem to be not following this trend. 
A least square fitting to the sample of SSCs excluding these 4 SSCs
gives us a relation of the form: $R_{\rm h}\propto M^b$ with
$b=0.29\pm0.05$. We used the orthogonal slope as defined in \citet{sixlinf}
to carry out the fitting. The observed power-law index is close to 1/3, the
value expected for tidally-limited clusters. 

The observed slope is distinct from 0.5, which is the value expected for virialized 
clusters \citep{Gieles2010}.  
Hence, at the outset it would seem that the majority of the M82 disk clusters are
tidally-limited, with the group of 4 massive clusters being more compact than their 
tidal values. We refer to this group of 4 clusters as the massive-compact 
SSCs henceforth in this paper. 
The tidal radius for a cluster of $10^5$~M$_\odot$ in the disk of M82 is 30--50~pc between galacto-centric
radius of 0.5~kpc and 4~kpc, which is much larger than the average 
half-mass radius at birth of 
compact clusters \cite[typically less than 1 pc,][]{Baumgardt2010,Banerjee2017}. Hence, clusters need to expand so as to be tidally-limited.
A detailed computation of the dynamical evolution of clusters under the tidal influence 
of M82's gravitational field is required in order to understand whether the majority
of M82 clusters are tidally-limited. We carried out such a study, whose results are
presented below.

\subsection{Cluster evolution using EMACSS tool}\label{subsec:emacss}

We carried out the evolution of M82 clusters using EMACSS \citep[Evolve Me 
a Cluster of StarS;][]{AlexanderGieles2012,AlexanderGieles2014}, 
a publicly available fast evolutionary code. 
In this simplified analytical code, 
the dynamical evolution of a cluster is treated in terms of the flow of 
energy normalised to the initial energy and relaxation time.
The code allows to study the evolution of clusters in the mass-radius plane
for an assumed initial mass-radius relation in the presence of a tidal field 
of a singular isothermal halo, which is parameterised by a flat rotation
curve \citep{Gieles2010}. 
Based on CO velocities for the nuclear region and
stellar and HI velocities in the rest of the disk \citet{Sofue1998}
suggested almost a Keplerian rotation curve for M82, implying the absence 
of a massive halo \citep[see also ][]{Mayya2009}. However, more recent studies
using the star cluster velocities suggest a flat rotation curve with a circular 
velocity of 100\,km\,s$^{-1}$ \citep[i.e.][]{Konst2009,Greco2012}.
In order to determine the influence of tidal forces in shaping the observed mass-radius 
relation at intermediate ages, we considered clusters at several values of galacto-centric 
radius ($R_{\rm g}$), covering the disk from 0.5 to 5 kpc, and evolved them from their birth 
up to 12~Gyr. The results are saved at specific epochs of interest, especially
at 100~Myr, which is the assumed canonical age of our sample of SSCs.

\begin{figure}
\begin{center}
\includegraphics[width= \columnwidth]{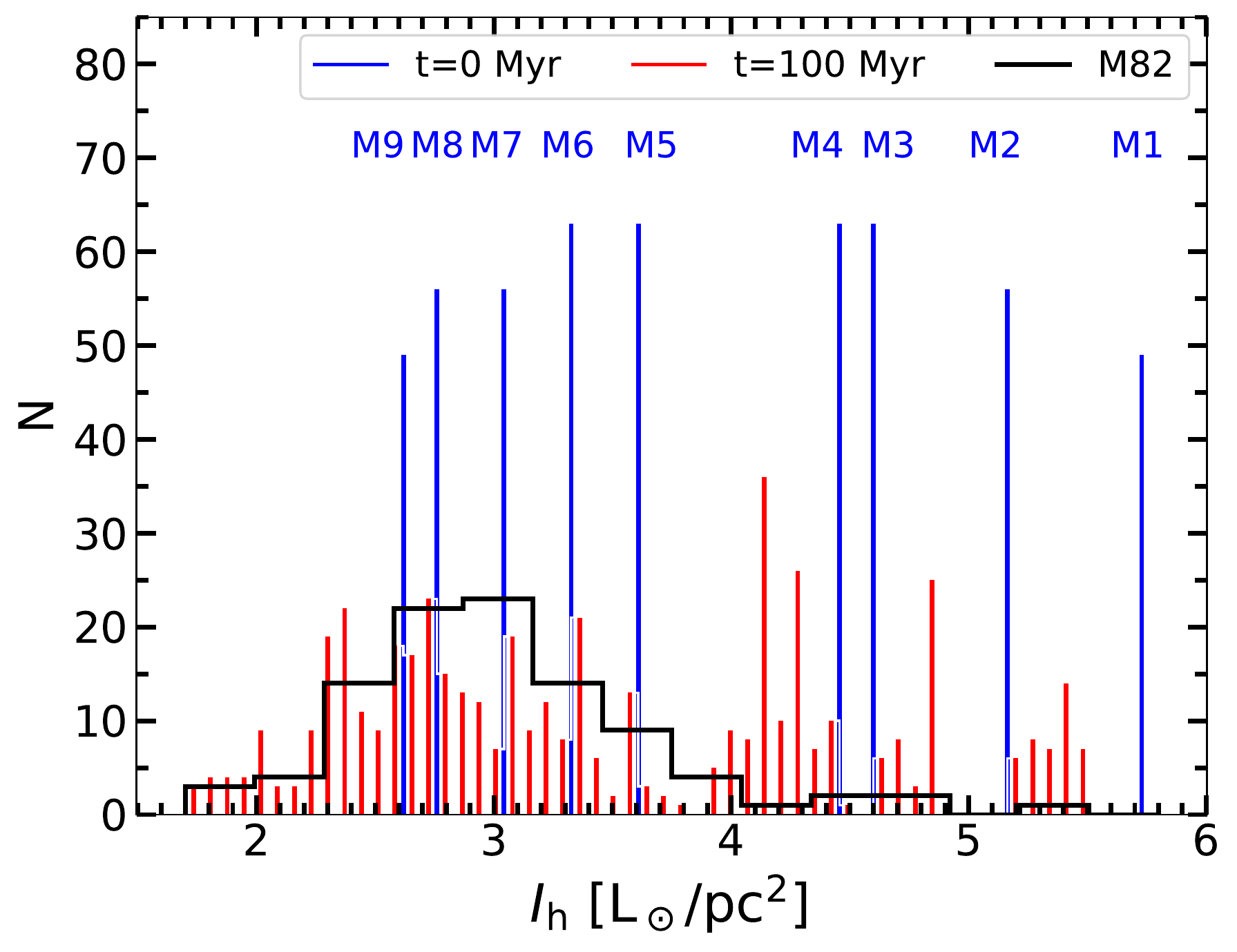}
\caption{
Comparison of the distributions of the mean surface brightness 
within the half-light radius. The observed distribution is shown by
the black histogram and the model values at 
t=0 and after 100~Myr of evolution
at a range of galacto-centric distances under M82's gravitational potential 
are shown by blue and red vertical bars, respectively. The height of each bar 
is the product of the number of mass-radius pairs used at each $I_{\rm h}$ and 
the number of discrete $R_{\rm g}$ values used. Mass-loss due to stellar evolution
and the expansion of the cluster lead to a decrease of $I_{\rm h}$ 
at 100~Myr. The $I_{\rm h}$ value at 100~Myr also depends on cluster's 
$R_{\rm g}$, which is the reason for multiple values of $I_{\rm h}$ at 100~Myr
for each model. 
The plot illustrates that the models
completely cover the observed range of mean surface brightness, with the 
majority of observed values lying in the range of $I_{\rm h}$ values at 100~Myr
for models M5 and less denser. 
}
\label{fig:Ih_hist}
\end{center}
\end{figure}

The first set of initial mass-radius relation we used is the virial mass-radius 
relation, following its use by \citet{Gieles2010}.
Under virial equilibrium, these relations correspond to constant surface
brightness within the half-mass radius ($I_{\rm h}$), in which case $R_{\rm h}\propto M^{0.5}$.
The complete set of initial mass-radius values explored by us is 
given in Tab. \ref{tab:emacss}, where each model name identified by M1, M2, 
M3, M4, M5, M6, M7, M8 and M9 corresponds to a particular mass-radius relation
with a fixed $I_{\rm h}$, and consists of a set of 9 values of mass and radius, 
identified by letters A, B, C, D, E, F, G, H, and K, with A and K
corresponding to the least and most massive cases in each relation, respectively.
Each model was evolved under the gravitational potential of M82 located at $R_{\rm g}$ 
values from 0.5~kpc to 5~kpc, which covers the full range of $R_{\rm g}$s for the disk SSCs in M82.
The distributions of initial (t=0) and present (t=100~Myr) $I_{\rm h}$ values are 
shown in Figure~\ref{fig:Ih_hist} by vertical bars of blue and red colours, respectively. 
The figure also shows the observed distribution
of $I_{\rm h}$ for our sample clusters, which is well within the range of values
covered by the 100~Myr models.

The four massive compact SSCs, which are outliers in Figure~\ref{fig:mass_vs_reff}, 
are well described by models M1 to M3. Evolution of these densest clusters
in the gravitational field of M82 follows the same locus as for isolated clusters, 
as can be inferred from the evolution of model M3 in the Appendix Figure~\ref{fig:ap_sim_emacss2}. 
Thus, the group of four massive clusters represents a subset of clusters
that evolve without the tidal influence of their host galaxy M82.
In other words, the evolution of these clusters is independent of 
their galacto-centric distances.

\begin{table*}
\caption{Models initial conditions}
\begin{center}
\begin{tabular}{|l|llllllllll|}
\hline
Point & $R_{h}$ & M1 & M2 & M3 & M4 & M5 & M6 & M7 & M8 & M9\\
 & (pc) & ($\log \frac{M}{M_\odot}$) & ($\log \frac{M}{M_\odot}$) & ($\log \frac{M}{M_\odot}$) & ($\log \frac{M}{M_\odot}$) & ($\log \frac{M}{M_\odot}$) & ($\log \frac{M}{M_\odot}$) & ($\log \frac{M}{M_\odot}$) & ($\log \frac{M}{M_\odot}$) &  ($\log \frac{M}{M_\odot}$)\\
 &  & ($I_h$=5.8)        & ($I_h$=5.2)   & ($I_h$=4.6)   & ($I_h$=4.5)   & ($I_h$=3.7)   & ($I_h$=3.4)   & ($I_h$=3.1)   & ($I_h$=2.8)   & ($I_h$=2.7)\\
(1) & (2) & (3) & (4) & (5) & (6) & (7) & (8) & (9) & (10) & (11)\\
\hline
A &      0.20   & 4.3   & 3.7   & 3.1   & 3.0   & 2.2   & 1.9  &  (X,X) & (X,X) & (X,X)\\
B &      0.38   & 4.8   & 4.2   & 3.7   & 3.5   & 2.7   & 2.4   & 2.1   & 1.8   & (X,X) \\
C &      0.74   & 5.4   & 4.8   & 4.3   & 4.1   & 3.3   & 3.0   & 2.7   & 2.4   & 2.3   \\
D &      1.41   & 6.0   & 5.4   & 4.8   & 4.7   & 3.9   & 3.6   & 3.3   & 3.0   & 2.9   \\
E &      2.71   & 6.5   & 5.9   & 5.4   & 5.2   & 4.4   & 4.1   & 3.8   & 3.5   & 3.4   \\
F &      5.21   & 7.1   & 6.5   & 6.0   & 5.8   & 5.0   & 4.7   & 4.4   & 4.1   & 4.0   \\
G &     10.00   & 7.7   & 7.1   & 6.5   & 6.4   & 5.6   & 5.3   & 5.0   & 4.7   & 4.6   \\
H &     (X,X)  & 8.2   & 7.6   & 7.1   & 6.9   & 6.1   & 5.8   & 5.5   & 5.2   & 5.1   \\
K &     (X,X) & (X,X)   & 8.2   & 7.7   & 7.5   & 6.7   & 6.4   & 6.1   & 5.8   & 5.7   \\
\hline
\end{tabular}
\hfill\parbox[t]{\textwidth}{Col (1):  Point designation with A having the smallest mass and radius, and with K the largest of each model line described in Cols (3)-(11).  Col (2) Initial half-mass radius $R_{\rm h}$ in parsecs.  Cols (3)-(11), logarithm of initial mass in units of $M_\odot$.   In each column is shown in parentheses the corresponding logarithmic surface brightness of each model in units of $L_\odot/pc^2$. (X,X) stands for clusters with inadmissible conditions (above or below the limits for the EMACSS code).}
\end{center}
\label{tab:emacss}
\end{table*}

\begin{table*}
\small\addtolength{\tabcolsep}{-5pt}
\caption{Evolved (100 Myr) to initial mass and radii ratios for models evolving under a tidal field and in isolation.}
\begin{center}
\begin{scriptsize}
\begin{tabular}{|l|llllllllll|}
\hline
Point & $R_{\rm g}$ & M1 & M2 & M3 & M4 & M5 & M6 & M7 & M8 & M9\\
 & (kpc)  & ($\rm q_{\it M}$,$\rm q_{\it R_{\rm h}}$,$R_{\rm hj_{0}}$) & ($\rm q_{\it M}$,$\rm q_{\it R_{\rm h}}$,$R_{\rm hj_{0}}$) & ($\rm q_{\it M}$,$\rm q_{\it R_{\rm h}}$,$R_{\rm hj_{0}}$) & ($\rm q_{\it M}$,$\rm q_{\it R_{\rm h}}$,$R_{\rm hj_{0}}$) & ($\rm q_{\it M}$,$\rm q_{\it R_{\rm h}}$,$R_{\rm hj_{0}}$) & ($\rm q_{\it M}$,$\rm q_{\it R_{\rm h}}$,$R_{\rm hj_{0}}$) & ($\rm q_{\it M}$,$\rm q_{\it R_{\rm h}}$,$R_{\rm hj_{0}}$) & ($\rm q_{\it M}$,$\rm q_{\it R_{\rm h}}$,$R_{\rm hj_{0}}$) & ($\rm q_{\it M}$,$\rm q_{\it R_{\rm h}}$,$R_{\rm hj_{0}}$) \\
(1) & (2) & (3) & (4) & (5) & (6) & (7) & (8) & (9) & (10) & (11)\\
\hline
A &     $\infty$        & (0.7,3.9,0)   & (0.7,5.1,0)   & (0.6,6.5,0)   & (0.6,6.8,0)   & (0.6,7.1,0)   & (0.7,5.0,0)   & (X,X,X)       & (X,X,X)       & (X,X,X)       \\
B &     $\infty$        & (0.8,1.6,0)   & (0.7,2.1,0)   & (0.7,2.8,0)   & (0.7,2.9,0)   & (0.7,3.8,0)   & (0.7,3.9,0)   & (0.7,3.8,0)   & (0.7,2.7,0)   & (X,X,X)       \\
C &     $\infty$        & (0.8,1.6,0)   & (0.8,1.6,0)   & (0.8,1.6,0)   & (0.8,1.6,0)   & (0.8,1.7,0)   & (0.8,1.9,0)   & (0.7,2.1,0)   & (0.7,2.2,0)   & (0.7,2.2,0)   \\
D &     $\infty$        & (0.8,1.4,0)   & (0.8,1.5,0)   & (0.8,1.5,0)   & (0.8,1.5,0)   & (0.8,1.6,0)   & (0.8,1.6,0)   & (0.8,1.6,0)   & (0.8,1.6,0)   & (0.8,1.6,0)   \\
E &     $\infty$        & (0.8,1.3,0)   & (0.8,1.3,0)   & (0.8,1.4,0)   & (0.8,1.4,0)   & (0.8,1.5,0)   & (0.8,1.5,0)   & (0.8,1.5,0)   & (0.8,1.5,0)   & (0.8,1.5,0)   \\
F &     $\infty$        & (0.8,1.3,0)   & (0.8,1.3,0)   & (0.8,1.3,0)   & (0.8,1.3,0)   & (0.8,1.3,0)   & (0.8,1.4,0)   & (0.8,1.4,0)   & (0.8,1.4,0)   & (0.8,1.4,0)   \\
G &     $\infty$        & (0.8,1.3,0)   & (0.8,1.3,0)   & (0.8,1.3,0)   & (0.8,1.3,0)   & (0.8,1.3,0)   & (0.8,1.3,0)   & (0.8,1.3,0)   & (0.8,1.3,0)   & (0.8,1.3,0)   \\
H &     $\infty$        & (X,X,X)       & (0.8,1.3,0)   & (0.8,1.3,0)   & (0.8,1.3,0)   & (0.8,1.3,0)   & (0.8,1.3,0)   & (0.8,1.3,0)   & (0.8,1.3,0)   & (0.8,1.3,0)   \\
K &     $\infty$        & (X,X,X)       & (X,X,X)       & (0.8,1.3,0)   & (0.8,1.3,0)   & (0.8,1.3,0)   & (0.8,1.3,0)   & (0.8,1.3,0)   & (0.8,1.3,0)   & (0.8,1.3,0)   \\
\hline
A &     2       & (0.7,3.9,0.01)        & (0.7,5.1,0.01)        & (0.6,5.9,0.02)        & (0.5,5.9,0.03)        & (0.4,4.4,0.05)        & (0.8,2.1,0.05)        & (X,X,X)       & (X,X,X)       & (X,X,X)       \\
B &     2       & (0.8,1.6,0.01)        & (0.7,2.1,0.02)        & (0.7,2.7,0.03)        & (0.7,2.8,0.03)        & (0.5,3.0,0.06)        & (0.5,2.7,0.08)        & (0.4,2.3,0.10)        & (0.8,1.2,0.10)        & (X,X,X)       \\
C &     2       & (0.8,1.6,0.01)        & (0.8,1.6,0.02)        & (0.8,1.6,0.03)        & (0.8,1.6,0.04)        & (0.7,1.6,0.07)        & (0.6,1.7,0.09)        & (0.6,1.6,0.12)        & (0.5,1.5,0.16)        & (0.5,1.4,0.17)        \\
D &     2       & (0.8,1.4,0.02)        & (0.8,1.5,0.03)        & (0.8,1.5,0.04)        & (0.8,1.5,0.04)        & (0.8,1.5,0.08)        & (0.8,1.5,0.11)        & (0.7,1.4,0.13)        & (0.7,1.4,0.17)        & (0.7,1.3,0.19)        \\
E &     2       & (0.8,1.3,0.02)        & (0.8,1.3,0.03)        & (0.8,1.4,0.05)        & (0.8,1.4,0.06)        & (0.8,1.4,0.10)        & (0.8,1.4,0.13)        & (0.8,1.4,0.16)        & (0.7,1.3,0.22)        & (0.6,1.2,0.24)        \\
F &     2       & (0.8,1.3,0.03)        & (0.8,1.3,0.04)        & (0.8,1.3,0.06)        & (0.8,1.3,0.07)        & (0.8,1.3,0.13)        & (0.8,1.3,0.16)        & (0.7,1.3,0.21)        & (0.6,1.0,0.28)        & (0.5,0.9,0.30)        \\
G &     2       & (0.8,1.3,0.03)        & (0.8,1.3,0.05)        & (0.8,1.3,0.08)        & (0.8,1.3,0.09)        & (0.8,1.3,0.16)        & (0.8,1.2,0.20)        & (0.6,1.0,0.27)        & (0.5,0.8,0.30)        & (0.4,0.7,0.30)        \\
H &     2       & (X,X,X)       & (0.8,1.3,0.06)        & (0.8,1.3,0.10)        & (0.8,1.3,0.11)        & (0.8,1.2,0.20)        & (0.7,1.0,0.26)        & (0.5,0.8,0.30)        & (0.3,0.5,0.30)        & (0.3,0.5,0.30)        \\
K &     2       & (X,X,X)       & (X,X,X)       & (0.8,1.3,0.12)        & (0.8,1.3,0.13)        & (0.7,1.0,0.26)        & (0.5,0.8,0.30)        & (0.3,0.6,0.30)        & (0.2,0.4,0.30)        & (0.1,0.3,0.30)        \\
\hline
\end{tabular}
\end{scriptsize}
\hfill\parbox[t]{\textwidth}{Col (1):  Point designation with A having the smallest mass and radius, and with K the largest of each model line described in Cols (3)-(11) in Tab. \ref{tab:emacss2}.  Col (2) Galactocentric radius $R_{\rm g}$ in kiloparsecs.  Cols (3)-(11), model mass at 100 Myr to initial mass ratio $\rm q_{\it M}$, model $R_{\rm h}$ at 100 Myr to initial $R_{\rm h}$ ratio $\rm q_{\it R_{\rm h}}$, model initial $R_{\rm h}$ to Jacobi radius ratio, ${R_{\rm hj_{0}}} = R_{\rm h,0}/R_{\rm j,0}$.  The latter values is 0 for isolated clusters, since in such cases ${R_{\rm j,0}}=\infty$. (X,X) stands for clusters with inadmissible conditions (above or below the limits for the EMACSS code).}
\end{center}
\label{tab:emacss2}
\end{table*}

\begin{figure*}
\begin{center}
\includegraphics[width= 0.85\textwidth]{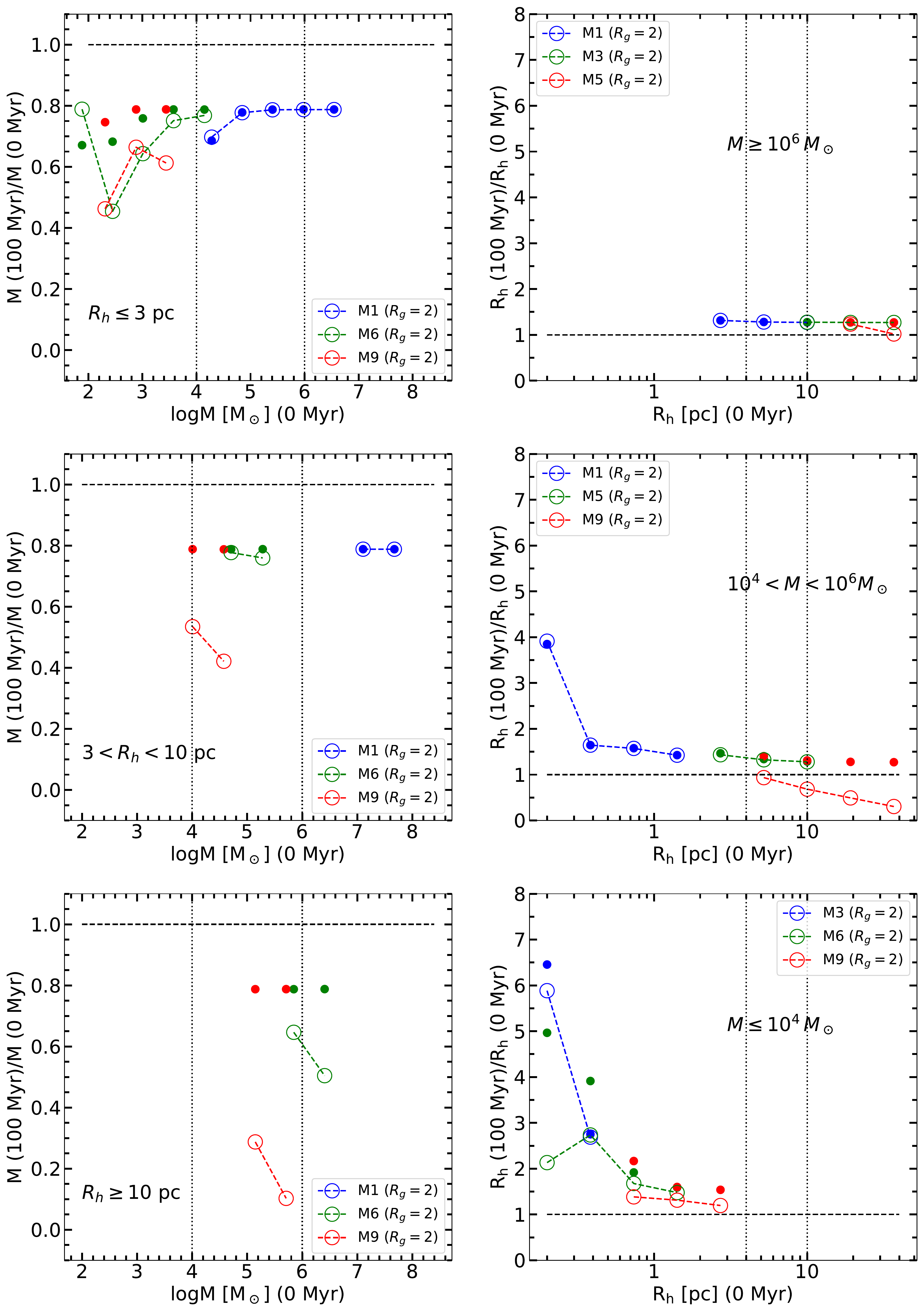}
\caption{
Theoretical ratio of evolved (100 Myr) to initial mass (left panels) and $R_{\rm h}$
(right panels) of SSCs in the disk of M82 using the cluster
evolutionary code EMACSS.
Mass ratios are shown for three ranges of radii, and
radius ratios are shown for three ranges of masses, each shown separately
in a panel for the sake of clarity. Dotted vertical lines show these
boundaries, and the dashed horizontal line shows unit ratio.
We carefully selected models (identified by M1, M6, M9 etc; see
Tab.~\ref{tab:emacss})
in each panel to illustrate all possible evolutionary scenarios.
Each cluster is evolved in isolation (empty circles) and under the
gravitational potential of M82 at $R_{\rm g}$=2~kpc (small filled circles).
See text for a detailed interpretation of the results.
}
\label{fig:rini_mini100}
\end{center}
\end{figure*}

\begin{figure}
\begin{center}
\includegraphics[width= 1.03\columnwidth]{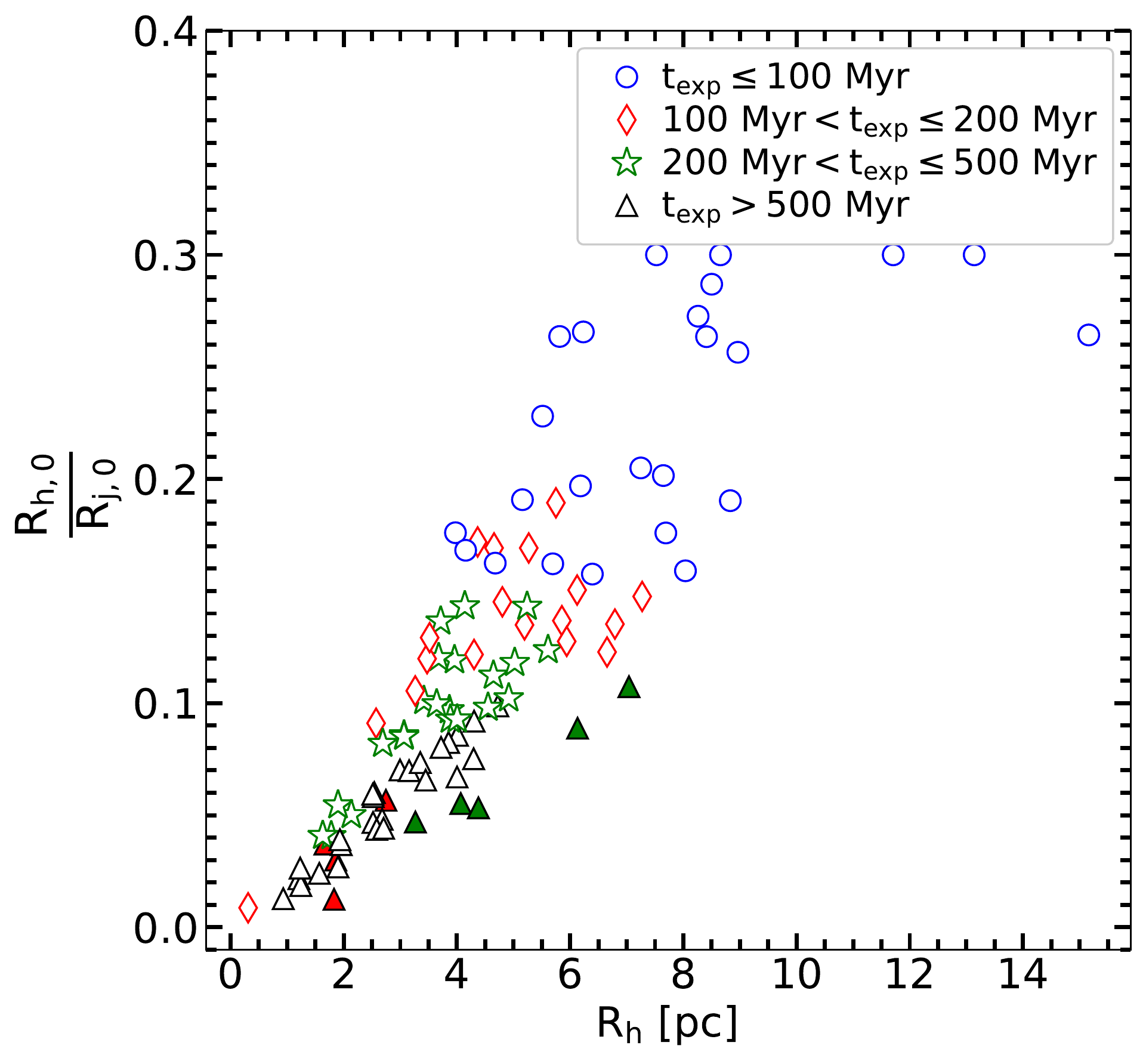}
\caption{Initial half-mass radius to Jacobi radius ratio $\frac{R_{\rm h,0}}{R_{\rm j,0}}$
vs half-mass radius at present coded according to the time taken to reach the maximum radius ($\rm t_{exp}$). 
The time bin corresponding to each symbol is shown in the figure inset. 
Note that for a given $\rm R_{\rm h}$, clusters with long $\rm t_{exp}$ have larger
$\frac{R_{\rm h,0}}{R_{\rm j,0}}$ and vice versa, with the tidally-limited clusters (blue circles) being the largest.
We show in red and green filled triangles the clusters surviving for a Hubble time, corresponding
to the massive-compact and outer-disk groups.}
\label{fig:rh_rjac}
\end{center}
\end{figure}

\begin{figure}
\begin{center}
\includegraphics[width= \columnwidth]{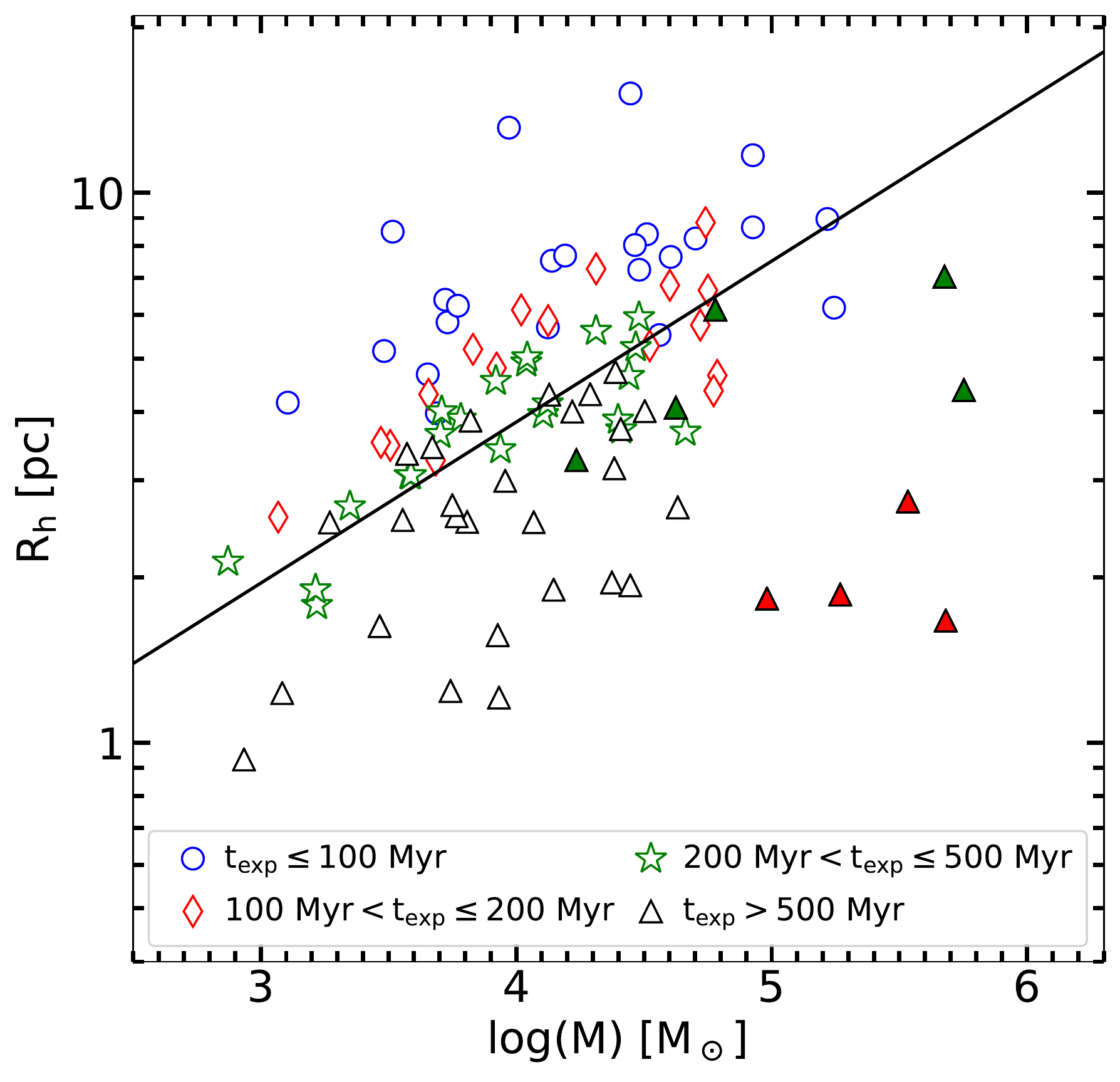}
\caption{Same as Figure~\ref{fig:mass_vs_reff}, but the symbols are
coded according to the time taken to reach the maximum radius ($\rm t_{exp}$). 
The time bin corresponding to each symbol is shown in the figure inset. 
Note that for a given mass, clusters with long $\rm t_{exp}$ have smaller
$\rm R_{\rm h}$ and vice versa, with the tidally-limited clusters (blue circles) being the largest.
We show in red and green filled triangles the clusters surviving for a Hubble time, corresponding
to the massive-compact and outer-disk groups.}
\label{fig:rh_vs_M_t_sr}
\end{center}
\end{figure}

In Figure~\ref{fig:Ih_hist}, it can be noticed that
the $I_{\rm h}$ values for majority of the clusters are below that for the model M3,
and lie between that for the models M4 and M9. 
We now explore the parameter-space of initial mass-radius values
that explain the observed trend for the majority of our clusters.

In Fig. \ref{fig:rini_mini100}, we compare the mass (left panels)
and the radius (right panels) of illustrative models at $t$=0 and $t$=100~Myr. 
The mass evolution is split into 3 radius bins, and radius evolution
is split into three mass bins. 
Given the mass-radius relation at birth, all models do not populate
all the diagrams. We have carefully chosen 3 models in each
panel to illustrate the overall behaviour of clusters in the first
100~Myr in the M82 disk. Numerical results for all models can be found in 
Table~\ref{tab:emacss2}. Location of models are shown both for
clusters evolving in isolation (solid circles) and under the potential
of M82 at $R_{\rm g}$=~2~kpc (empty circles joined by dashed lines).  
Models for which evolutionary results at 2~kpc is identical to evolution
in isolated conditions are those for which locations of empty circles,
which are intentionally shown bigger, coincide with that of solid circles. 
Evolution of these models would be identical at all galacto-centric
distances larger than 2~kpc. 
All clusters are in a state of expansion during the first 100~Myr for 
all of our models. In the presence of tidal fields, the expansion is
halted when the cluster radius reaches the tidal radius.

We first discuss the mass-loss during the first 100~Myr of evolution.
All clusters loose a minimum of 20\% of mass, which is due to mass-loss
during stellar evolution, rather than dynamical processes.
Compact clusters ($R_{\rm h}<$3~pc) do not loose additional mass if they
are more massive than $10^4$~M$_\odot$. On the other hand, 
clusters born with ($R_{\rm h}>$3~pc) are all susceptible to mass-loss
due to dynamical effects,
unless they are more massive than $\sim10^7$~M$_\odot$.
Clusters born with average surface brightness $\log(I_h)<$3.4~L$_\odot/{\rm pc}^2$
(models M6--M9) loose more than 30\% of their initial mass.
As expected, the quantity of mass lost is larger if they are born
less dense and/or with a large radius.

The plots involving radius evolution
suggest that all massive SSCs ($M>10^6$~M$_\odot$) 
denser than M5 evolve identical to that of isolated clusters
and are in a state of expansion. Clusters born with surface densities
similar or less than that of M5 have already expanded to their
maximum radius and are tidally-limited.
For the intermediate mass range ($10^4<M<10^6$~M$_\odot$), 
only very dense (denser than M5) and those born with small radius 
($R_{\rm h}\lesssim$4~pc) are not perturbed by the tidal fields at 2~kpc
(empty circles and solid circles coincide), and
hence they are in a state of expansion. 
Rest of the models have already tidally-limited (i.e. empty 
circles are below the solid circles).
The trend is similar for $M<10^4$~M$_\odot$, except that clusters
have to be born denser than M4 model and $R_{\rm h}\lesssim$1~pc 
so as to avoid expanding to the tidal limit at $R_{\rm g}$=2~kpc. 
Model M9 represents a special set of initial conditions, characterised
by initial radius larger than the tidal radius at $R_{\rm g}$=5~kpc. 
Clusters with these initial conditions do not survive at smaller $R_{\rm g}$.

The evolutionary trajectory at other values of $R_{\rm g}$ is similar to that described 
above. At $R_{\rm g}<2$~kpc, majority of the models less dense than M6 are
tidally-limited at 100~Myr of evolution. This is illustrated in
Fig.~\ref{fig:ap_sim_emacss2} (in the Appendix). 
At $R_{\rm g}>2$~kpc, the behaviour is similar to that of isolated models.

\begin{figure*}
\begin{center}
\includegraphics[width= \textwidth]{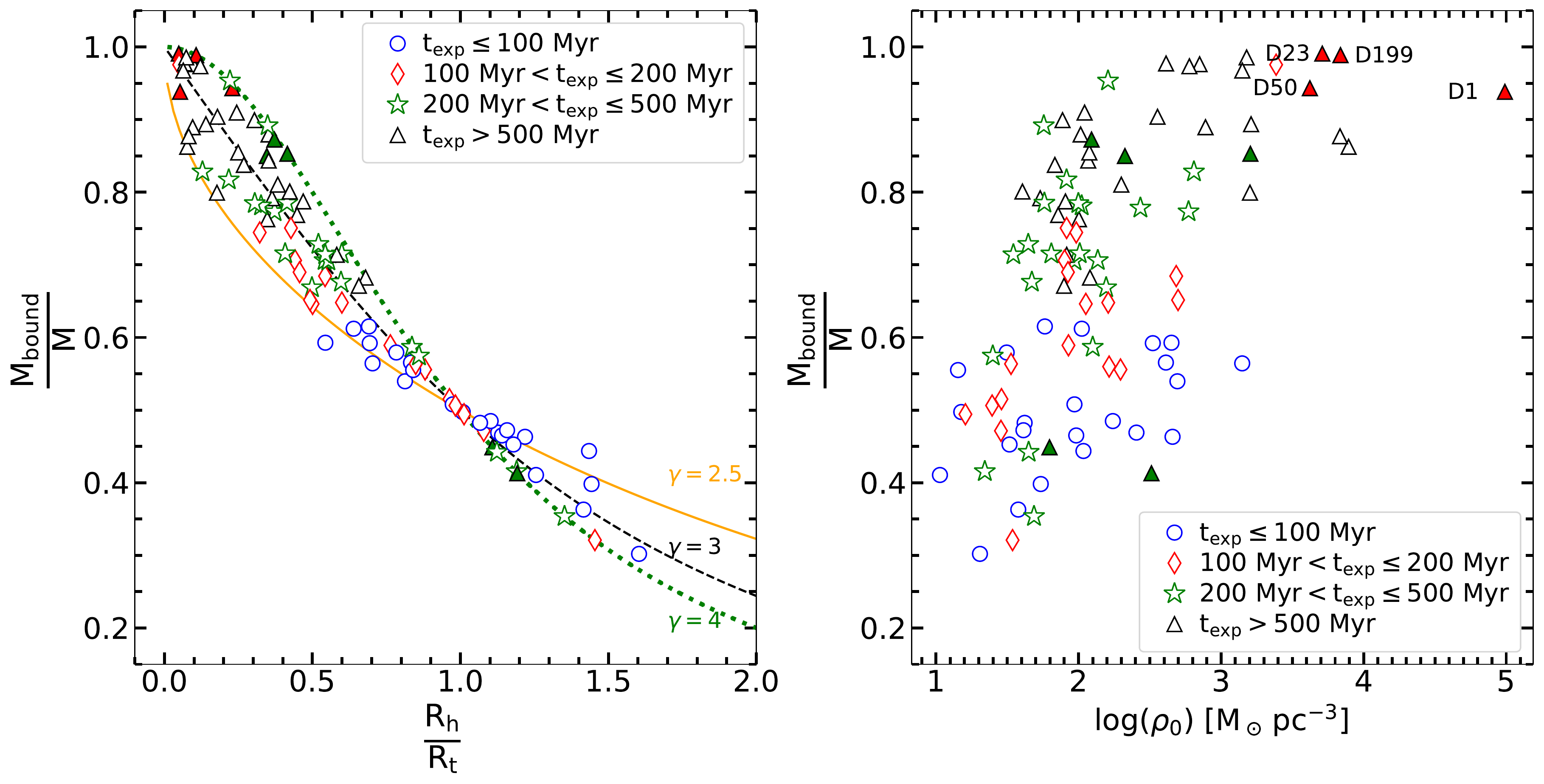}
\caption{
(Left) Bound mass to total cluster mass ratio ($\frac{M_{\rm bound}}{M}$) vs 
half-light radius to tidal radius ratio ($\frac{R_{\rm h}}{R_{\rm t}}$), coded 
in terms of $\rm t_{exp}$, following the color codes indicated in the inset.
The solid, dotted and dashed lines, represent the bound mass to total cluster 
mass ratio for given Moffat-EFF index and $\frac{R_{\rm h}}{R_{\rm t}}$ ratio.
(Right) $\frac{M_{\rm bound}}{M}$ versus the central mass volume density ($\rho_0$). 
The SSCs surviving for the Hubble time are highlighted by filled symbols
(red triangles: four massive-compact SSCs; green circles: five compact outer-disk SSCs). 
The IDs of the massive-compact SSCs are indicated.}
\label{fig:rh_rt_vs_Mbound_M}
\end{center}
\end{figure*}

In summary, cluster evolution critically depends on the initial mean stellar 
density of the clusters, which is related to the observed
mean surface brightness $I_{\rm h}$ through the mass-to-light ratio, $\Gamma$. 
Clusters expand during this phase, hence $I_{\rm h}$ decreases. 
The amount of change in $I_{\rm h}$ depends on the initial $I_{\rm h}$, with models less denser than
M4 forming a broad group with $\log(I_{\rm h})$={2.4--3.6}~$\rm L_\odot/pc^2$,
which covers the observed range of $I_{\rm h}$ of M82 SSCs. 
At the same time, the three highest density models 
(M1, M2 and M3) separate from this group. The group of four massive-compact
SSCs has similar range of surface densities as these models.
The plot illustrates that models M1, M2, M3 represent the condition of
the group of the four massive-compact SSCs, whereas the models M4--M9 represent
that of the rest of the M82 disk SSCs, which are all in the rapid expansion phase.

Having discussed the evolutionary behaviour of clusters in the disk of M82,
we now return to the implications of the observed mass-radius relation for SSCs
in the disk of M82 in Figure~\ref{fig:mass_vs_reff}. In particular, we address 
the question of whether the derived power-law slope $b=0.29\pm0.05$, which is 
close to the expected value for tidally bound clusters, does really imply the M82 SSCs are tidally bound?
In order to address this question, we evolve each M82 SSC using EMACSS so as to
reproduce the currently observed $M$ and $R_{\rm h}$ at t=100~Myr at their currently observed $R_{\rm g}$.
This was achieved by evolving clusters for a variety of initial guess values of $M$ and $R_{\rm h}$, all placed
at the currently observed $R_{\rm g}$ of the SSC in analysis.
In particular, we follow the evolution from t=0 up to the time when the expanding
cluster fills the entire volume defined by its tidal radius. 
In the EMACSS context, the proxy of a cluster volume filling is the
half-mass radius to Jacobi Radius ($R_{\rm j}$) ratio $\frac{R_{\rm h}}{R_{\rm j}}$ 
\citep{AlexanderGieles2014}, with $\frac{R_{\rm h}}{R_{\rm j}}$ being greater than 0.19 
for tidally limited clusters and less than 0.1 for clusters completely embedded within their tidal radius.
A cluster with $\frac{R_{\rm h}}{R_{\rm j}}>0.19$ starts loosing its stars leading to a decrease in its mass and size.
Time taken to reach this radius is directly proportional to the initial mean surface density of the clusters.
In Figure~\ref{fig:rh_rjac}, we plot  the initial values of $\frac{R_{\rm h}}{R_{\rm j}}$ in terms 
of their current half-mass radii.  Points are coded with coloured symbols that are indicative of the time
 taken to reach the maximum radius, or the expansion timescale, $t_{\rm exp}$.  The majority of 
 tidally limited clusters (blue circles) occupy the upper-envelop of the diagram, with values above
 0.17, close to the limiting value 0.19 representing clusters filling their tidal volume.
 Interestingly, these clusters are on average larger (>4 pc) than clusters with lower
  $\frac{R_{\rm h,0}}{R_{\rm j,0}}$ values, suggesting that the present-day tidally-limited clusters
  have been born large, filling a large fraction of their tidal volume. On the other hand, clusters
 filling their tidal volume at times larger than 500~Myr, are compact (<4 pc) and have 
 $\frac{R_{\rm h,0}}{R_{\rm j,0}}$ values lower than 0.1, suggesting that compact clusters have been born
 well-embedded within their tidal volume.
In Figure~\ref{fig:rh_vs_M_t_sr}, we replot the mass-radius relation, coding each 
point as in the case of Figure~\ref{fig:rh_rjac}, in terms of $t_{\rm exp}$.
Twenty three clusters, all with $R_{\rm h}>$3~pc, that occupy the top-most envelope of the diagram
(blue circles) are tidally limited. 
The rest of the clusters ($R_{\rm h}\leq$3~pc)
are still expanding at their present age of 100~Myr, and are not yet tidally-limited
in spite of following the relation expected for the tidally-limited clusters. 
Thirty six of these expanding clusters would take more than 500~Myr to reach the
tidal limit, with the four compact-massive clusters taking more than 5~Gyr to do so.
In general, low-mass clusters expand at a greater rate as compared to the massive
clusters which leads to flattening of the mass-radius relation.
Thus, clusters that
are born with Virial equilibium would have slope $b<$0.5, when they are in the expansion phase.
This is clearly seen in Figure~\ref{fig:ap_sim_emacss2} (the bottom-panel), where
the trajectory of models M6 at $R_{\rm g}$=0.5~kpc at 100~Myr (dashed blue line)
is flatter than the initial trajectory (dashed black line; slope=0.5) and is nearly
parallel to the observed relation (black line).
This explains the observed slope of $b=0.29$ for the M82 disk clusters.

In EMACSS, the cluster mass is contained within a finite radius. Hundred percent of
the cluster stars are bound as long as this radius is smaller than the tidal radius.
These characteristics are satisfied by the King and Plummer profiles used by EMACCS 
for modelling the effects of the tidal field. 
However, Moffat-EFF profiles have finite mass, but
over an infinite radius. Hence, there is always some amount of mass outside the tidal radius. 
These stars belong to an unbound halo. 
For example, for a typical cluster of $\gamma=2.7$, 30\% and 20\% of the total 
cluster mass is in unbound stars for $R_{\rm h}/R_{\rm t}$=0.6 and 0.4, respectively. 
As argued in Section~\ref{subsec:pars}, it takes typically 1~Gyr, which is
10 times the age of our clusters, to get rid of these unbound stars. 
This is the reason why Moffat-EFF profiles are better fits than the King 
profiles for the M82 disk SSCs \citep{Cuevas2020}.

The mass of our clusters is derived by multiplying the total luminosity of the
fitted Moffat-EFF profile by the mass-to-light ratio. Thus our derived mass
includes the stellar mass in unbound halos.
The calculation of the tidal radius for each M82 SSC allows us to determine the
fraction of the bound mass, which is the integration of the Moffat-EFF mass profile
up to $R_{\rm t}$, to the total mass (Eq. \ref{eqn:Mbound}). This fraction is expected to decrease
for clusters occupying a larger fraction of their tidal radius. 
In Figure~\ref{fig:rh_rt_vs_Mbound_M} (left), we show this dependence,
where we code the symbols based on their expansion time scale. Tidally-bound
clusters and clusters in an advanced stage of expansion have less than
80\% of their total mass in bound stars. 
Dense clusters have a higher fraction of mass in bound stars, which
is shown in the Figure on the right.

\begin{figure*}
\begin{center}
\includegraphics[width= 0.9\textwidth]{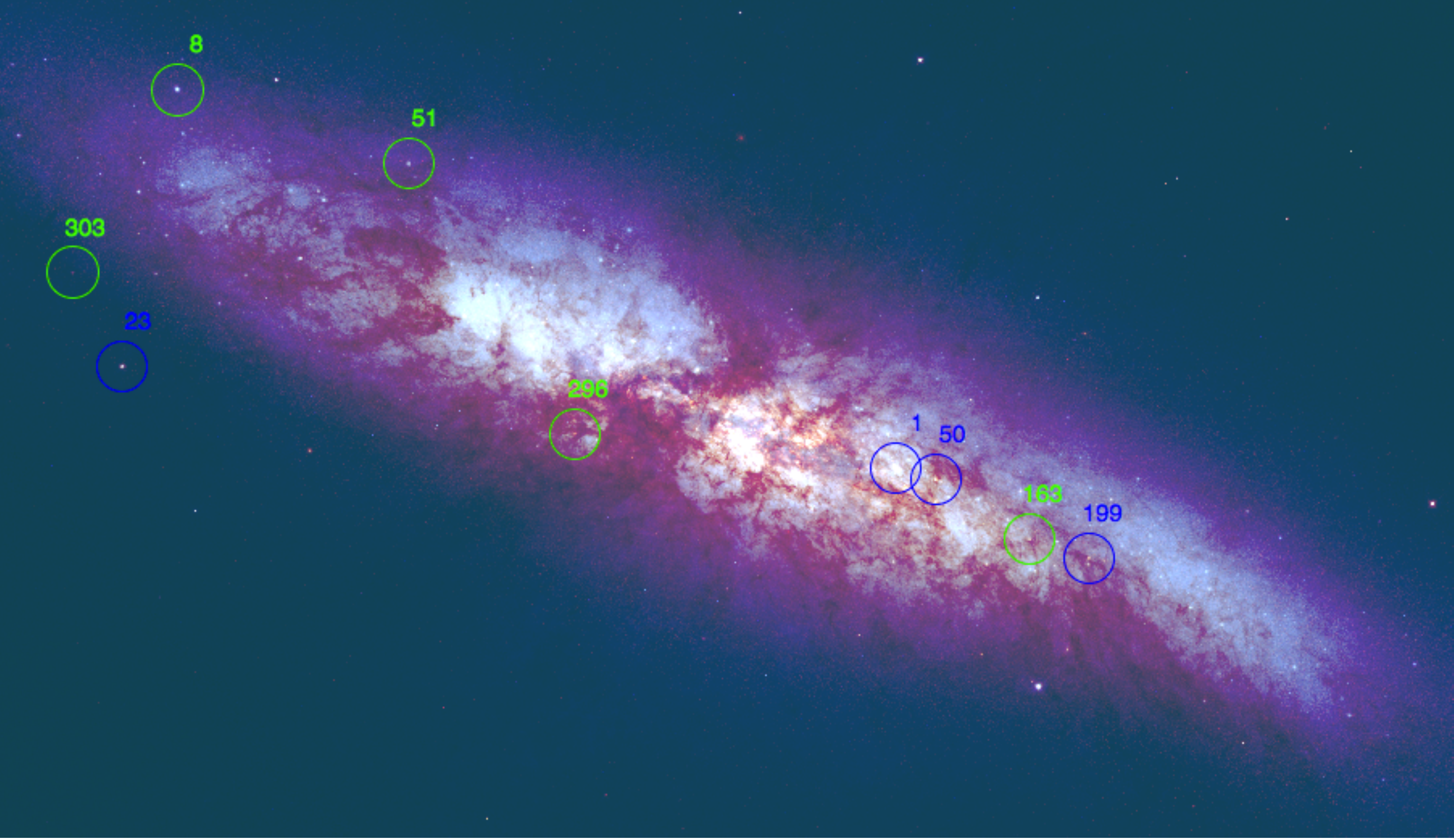}
\caption{RGB image of M82 formed using HST images in filters F435W,
F555W and F814W as blue, green and red components, respectively.
The groups of four massive-compact SSCs and 
five compact outer-disk SSCs,
that survive for Hubble time, are identified by blue and green circles,
respectively.}
\label{fig:ap_m82}
\end{center}
\end{figure*}

\begin{figure}
\begin{center}
\includegraphics[width= \columnwidth]{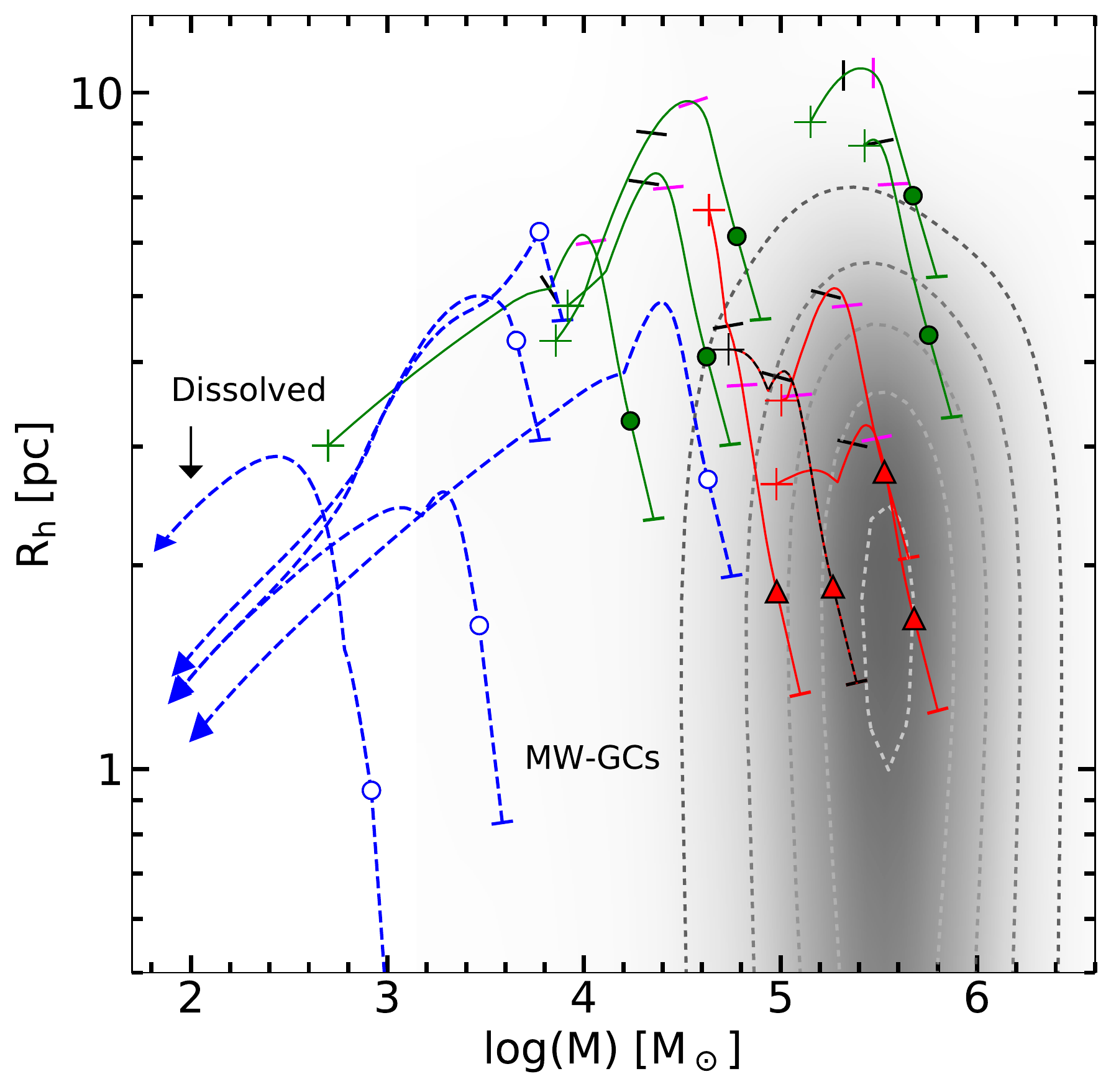}
\caption{Long-term evolution 
in the mass-radius plane of illustrative M82 disk SSCs. 
The solid curves correspond to the two groups of clusters (massive-compact SSCs, shown in red 
and compact outer-disk SSCs, shown in green) that
survive for the Hubble time, whereas the dashed curves represent  
clusters that dissolve well before the Hubble time.
Evolution begins (t=0) at the bottom-right corner of the track (shown by a tick mark),
passes through the current location (shown by triangles or circles) and ends 
at the plus symbol for surviving clusters and an arrow for dissolving clusters.
Tick marks along the track are shown at 500~Myr and 2~Gyr, for clusters that are
still surviving at these ages
(two massive-compact SSCs follow identical tracks, and hence for the sake of
visualization, we have shown one of them with a black line).
The location of the Milky Way Globular Clusters in this diagram,
extracted from \citet{Harris1996}, is shown in gray-scale, with the 
contours marking the boundaries that enclose 86\% (outer-most), 70\%, 
61\%, 41\% and 15\% (inner-most) of the GCs.
The darker zones indicate higher number density of GCs in this diagram.
The end-point of the evolutionary locus of massive-compact 
SSCs is in the range of observed values of Galactic GCs, suggesting these 
SSCs are proto-GC candidates.
}
\label{fig:mass_vs_reff_12Gyr}
\end{center}
\end{figure}

\subsection{Long-term evolution and globular cluster progenitors in M82}\label{Sec:long}

In the previous section, we established that the majority of the clusters in the disk 
of M82 are in expansion at present and around half of those would take more than 500~Myr
to start experiencing the disrupting effect of tidal forces.
In order to establish whether any of these SSCs would survive for Hubble time to 
become GCs,
we evolved each of our sample SSCs up to an age of 12 Gyr, using EMACSS 
from their present-day parameter values. EMACSS is run keeping each SSC at
its currently observed galacto-centric distance. We find that only $\sim$~9\% 
of the sample (9 SSCs; shown by filled symbols in Figure~\ref{fig:rh_vs_M_t_sr}) 
would remain bound after 12~Gyr evolution. These include
all the four massive compact SSCs (D1, D23, D50, D199), and a group
of five SSCs (D8, D51, D163, D296, D303), which we called compact 
outer-disk SSCs. These latter group of SSCs survive due to their large 
galacto-centric distances ($R_{\rm g}>2$~kpc). The rest of the SSCs 
do not survive for Hubble time.
The locations of these nine surviving SSCs are marked in an RGB image in 
Figure~\ref{fig:ap_m82}. Paper I contains plots of surface 
brightness profiles and individual RGB images of all the 99 SSCs,
including these nine clusters.

How do their final mass and radius compare with those of the Galactic GCs?. In order to
answer this question, in Fig. \ref{fig:mass_vs_reff_12Gyr} we show the
evolutionary locus in mass-radius diagram for the nine surviving SSCs.
For comparison, we also show the evolution of five dissolved clusters.
Red and green lines indicate SSCs belonging to the massive-compact 
and compact outer-disk groups, both of which survive for Hubble time.
The blue line indicates five illustrative dissolving SSCs. 
The final positions of these SSCs are indicated by plus symbols
for surviving clusters and by arrows for dissolving clusters.
We also show in gray-scale with overlaid contours the distribution of the
Galactic GCs in this diagram. Our group of massive-compact clusters
ends up with similar masses, but with larger radius than the Galactic GCs, according
to the EMACSS predictions. We note that the final size of the real clusters
is controlled by the core-collapse, whose treatment in EMACSS is only
simplistic and is not reliable at times 
much longer than the typical core-collapse timescale.
N-body simulations that includes a realistic 
treatment of core-collapse would be required to predict the final 
post core-collpase radius of these clusters. 
The final radius obtained by EMACSS at best could be considered as an upper limit.
The final obtained mass, on the other hand, is reliable. The location of the 
surviving massive-compact SSCs in this diagram suggests four SSCs would 
have masses slightly lower than the median mass for the Galactic GCs ($\sim3\times10^5$~M$_\odot$).
The fourth one occupies the low-mass end of the Galactic GC mass distribution.
The group of five surviving outer-disk SSCs ends up with 
systematically larger radii, and slightly lower masses, as compared to the group of 
massive-compact SSCs,
and hence these are unlikely to be GC-progenitors.
Their late-time characteristics resemble very much the
characteristics of faint fuzzies discovered by \citet{Larsen2000}. We
will discuss more about this in \S\ref{sec:fundplane}. 

We draw special attention to the widely-studied cluster M82-F 
(our cluster D1), the most massive SSC of the group of four surviving massive-compact clusters. 
This cluster had been discussed by \citet{SmithGallagher2001} as a doomed cluster.
Their conclusion was not based on dynamical grounds, instead based on the 
apparent lack of long-living low-mass stars, which they inferred from the
peculiarly low mass-to-light ratio for this object. With the top-heavy IMF, M82-F
would not have any stars in the main sequence after $\sim$2~Gyr of age 
\citep{SmithGallagher2001,McCrady2005,deGrijs2007rev}.
On the other hand, we assumed a standard Kroupa IMF, which has sufficient low-mass
stars to keep the cluster dynamically stable over the Hubble time.
  
EMACSS code does not include two processes that are known to play principal role
in disrupting a cluster, especially in the disk of a galaxy. These are disk shocks
experienced due to the presence of non-axisymmetric structures such as a bar and 
spiral arms, and disruption caused due to the interaction of the clusters with
Giant Molecular Clouds (GMCs) during their passage through spiral arms.
In the first place, spiral arm passages and disk shocks act in a similar way, 
driving a comparable mass-loss to that of the influence of tidal fields 
\citep{LamersGieles,Lamers2010}. Moreover, it is well-known that at least two 
episodes of disk shocks are required to disrupt a cluster \citep{VesperinoHeggie1997}.
These processes affect severely for cluster masses below $10^4$~$M_\odot$, being
the most critical one the encounter with GMCs. The disruption time for a cluster with a mass of $10^4$ $\rm M_\odot$, $R_{\rm h}=$3.75~pc interacting with a Milky Way-like GMC (with $\rho_n=0.03\, \rm M_\odot pc^{-3}$, $\Sigma_n=170\, \rm M_\odot pc^{-2}$,                $\rm \sigma_{cn}=\sqrt{\sigma_n^2+\sigma_c^2}=10$~km/s, with $\rm \sigma_n$, and $\rm \sigma_c$, the cloud and cluster velocity dispersions, respectively)
is of $\sim$~2 Gyr \citep{Gieles2006}.  From Eq. 22 in \citet{Gieles2006} 
the disruption due to encounters with Milky Way-like GMCs can be re-written as

\begin{equation} \label{eq:fact_GMC}
\begin{split}
t_{dis}  =  2.43 &\bigg{(}\frac{\eta}{0.4}\bigg{)}\bigg{(}\frac{0.25}{f}\bigg{)}\bigg{(}\frac{2.5}{g}\bigg{)}\bigg{(}\frac{\sigma_{\rm cn}}{10\, \rm{\frac{km}{s}}}\bigg{)}\bigg{(}\frac{5.1\, {\rm M_\odot^2} {\rm pc}^{-5}}{\Sigma_n \rho_n}\bigg{)}\\
& \bigg{(}\frac{R_{\rm h}^2/\overline{R}^2}{0.67}\bigg{)}\bigg{(}\frac{M_{\rm c}}{10^4\, {\rm M_\odot}}\bigg{)}\bigg{(}\frac{3.75\, {\rm pc}}{R_{\rm h}}\bigg{)}^3\, {\rm Gyr}
\end{split}
\end{equation}
\noindent with $f$, $g$, $\eta$, and  $R_{\rm h}^2/\overline{R}^2$, dynamical parameters dependent on the environment and dynamical state of the cluster and the cloud (for further details refer to \citet{Gieles2006}). From the latter expression, we obtain $t_{dis}\sim$~62 Gyr for D23, the least massive cluster of the massive-compact group,
which is a factor of $\sim$25 larger than the expected $t_{dis}$ for the canonical cluster defined by \citet{Gieles2006}. This large disruption time is due to its 10 times larger mass and two times smaller radius as compared to that of the canonical cluster. The disruption time due to collisions with the Milky Way-like GMCs, in general, is larger than the Hubble time for clusters more  massive than $10^5$ $\rm M_\odot$ and more compact than 3 pc. The previously stressed arguments suggest that the four massive compact SSCs are good proto-GC candidates. On the other hand, we find that our fuzzy cluster candidates are prone to be disrupted by GMC encounters in $\sim$~1~Gyr.

\subsection{$\mu_0$ vs $R_{\rm c}$ scaling relation}\label{sec:fundplane}

Most of the Galactic GCs fall on a straight line in log-log plots formed by parameters
such as the central surface brightness ($\mu_0$) and core radius ($R_{\rm c}$).  
This is attributed to the existence of a fundamental plane for these old stellar 
systems \citep{Djorgovski1995,McLaughlin2000}. It is not clear whether systems 
of only a few hundreds of million years belong to this fundamental plane.    
We discuss below a projection of this diagram for M82 clusters.

\begin{figure}
\begin{center}
\includegraphics[width= \columnwidth]{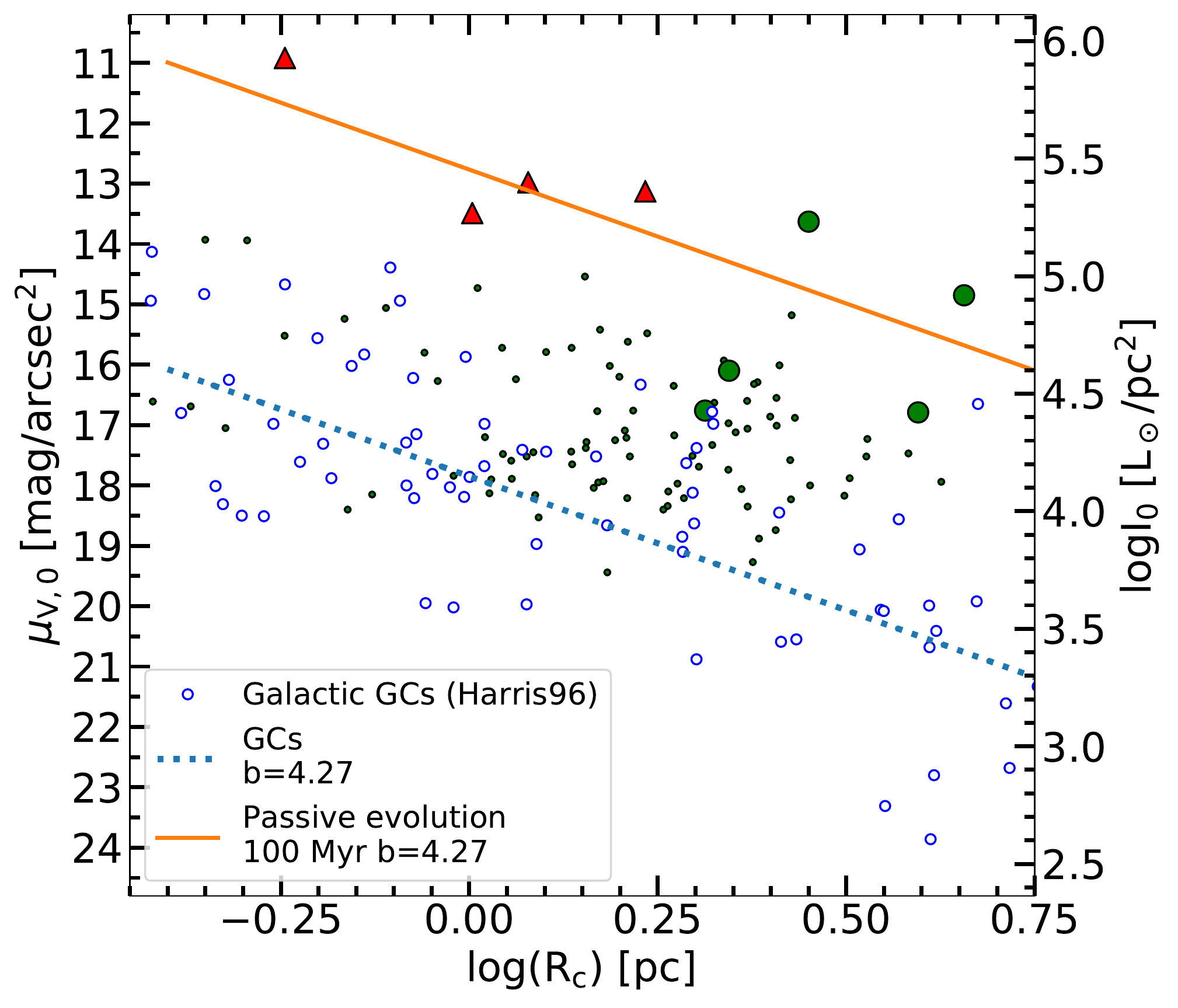}
\caption{
Central surface brightness $\mu_0$ vs  core radius $R_{\rm c}$ diagram for
M82 disk SSCs.
The $\mu_0$ vs  $R_{\rm c}$ scaling relation for GCs, is plotted 
with a blue dashed line with slope b=4.27, which is obtained
by fitting a straight line using the least-squares method to the Galactic GCs sample
of \citet{Harris1996} (blue empty circles). The
orange solid line corresponds to the passive evolution of GCs (going backwards
to an age of 100 Myr). The groups of four massive-compact, and five compact
outer-disk SSCs that survive for Hubble time are identified by red filled triangles
and green solid circles, respectively.
The figure illustrates that the current properties of the former group
are consistent with these being proto-GCs, 
whereas the SSCs of the latter group systematically end up 
fainter in $\mu_0$ and larger in $R_{\rm c}$ as compared to the Galactic GCs.
}
\label{fig:mu_vs_core}
\end{center}
\end{figure}

In Fig.~\ref{fig:mu_vs_core}, we plot $\mu_0$ in the $V$-band against $\log$($R_{\rm c}$) 
for M82 SSCs, and Galactic GCs. The two groups of SSCs 
surviving for Hubble time are shown by
red filled triangles (massive-compact clusters) and green solid circles 
(compact outer-disk clusters). The rest of the SSCs are shown
by black dots, and the Galactic GCs, extracted from  
\cite{Harris1996} by blue empty circles.
We fitted a straight line to the latter sample using a least-square fit
which is shown by a blue dotted line, with slope $b=4.27\pm0.30$.
It can be observed that both the SSCs and the GC samples 
have large dispersions, with the GC sample dispersion 
(2.77 mag arcsec$^{-2}$) twice larger
than that of SSCs (1.45 mag arcsec$^{-2}$).
The surviving SSCs fall on a sequence which is parallel to the
GC relation, but is shifted to brighter $\mu_0$.

This difference might be explained by two evolutionary effects: 
dynamical evolution of clusters that can change both $\mu_0$ and $R_{\rm c}$
and passive evolution leading to an increase in mass-to-light ratios. 
The former effect is related to the core-collapse, which is expected to
maintain the core luminosity constant \citep{Kormendy1985,Kuper2008}. 
Under these circumstances, $I_0\propto R_{\rm c}^{-2}$, i.e.
the SSCs will move along a line of slope 5 in Fig.~\ref{fig:mu_vs_core}, which is only slightly steeper 
than the plotted line ($b=4.27$). Thus, due to core-collapse
the intercept of the line defined by our group of massive-compact SSCs would not change much. 
If all the sample SSCs have 
the same age of 100~Myr, our clusters are expected to be 5~mag fainter 
in $\mu_0$(V)\footnote{These mass-to-light ratios take into account the 
mass lost due to stellar evolution.} due to passive evolution \citep{BruzualCharlot2003}. 
We plot in orange the line corresponding to passive evolution of GCs (evolving 
backwards from the GC relation), for a uniform age of 100 Myr.  It can be 
noticed that the four massive-compact SSCs lie on this line. 
Thus, this group of four massive-compact SSCs not only survive, but also would 
occupy the same fundamental plane as for the Galactic GCs.

Two of the five SSCs previously discussed as compact outer-disk clusters, 
lie close to the orange line, with the remaining 3 being fainter by $\sim$2--2.5~mag. 
These clusters after undergoing evolution upto $>$12~Gyr would be fainter and bigger than 
the present-day GCs in the Milky Way, thus are unlikely to be classified as GCs. 
On the other hand, these are excellent candidates for the red extended clusters, also known as
faint fuzzies, seen in M101 \citep{Simanton2015}, and lenticular galaxies 
\citep{Larsen2000}. In the lenticular galaxy NGC\,1023, the faint fuzzies
belong to the disk \citep{Chiessantos2013} and are found in a ring at a galacto-centric radius 
of $\sim$5~kpc \citep{LarsenBrodie2000,BrodieLarsen2002}, very similar to the characteristics we find for the
group of compact outer-disk SSCs.

\section{Conclusions}\label{Sec:Conclusions}

In this work, we present the complete set of structural parameters
corresponding to Moffat-EFF profiles for a sample of 99 SSCs in the 
disk of M82. The sample had been earlier analysed in Paper~I and
makes use of HST/ACS imaging data in F435W, F555W and F814W bands.
The quantities presented in this work are: core and half-light
radius, tidal radius, central surface and mass densities, mean surface density at half-light
radius, total and bound mass, luminosities, and projected central velocity dispersions.
The fact that the SSCs in the disk of M82 were born in a disk-wide 
burst around 100--300~Myr ago, allows us to address the evolutionary
behaviour of a relatively large and homogenous sample of
clusters of intermediate ages. In particular, we 
discuss the mass function, the size function, and the mass-radius 
relation in our sample of SSCs, and compare these with similar
data in other galaxies.
The mass distribution follows a power-law function of index $\alpha$=1.5
for masses above $10^4$~M$_\odot$, a result similar to that
obtained for the full sample of 393 SSCs by \citet{Mayyacat}. 
This index is flatter than that found for young SSCs ($\alpha$=2.0).
On the other hand, the distribution of half-light radius follows 
a log-normal form centered at 4.26~pc. We compare the distributions
of mass and $R_{\rm h}$ for M82 with existing data for similar-age (50--500~Myr)
and older ($>500$~Myr) SSCs in other galaxies. We find that the M82 
mass and $R_{\rm h}$ distributions agree very well with those in the LMC/SMC
for similar-age clusters. The distributions also 
compare well with those in the giant spiral galaxy M83, but for 
slightly older clusters. On the other hand, the distributions
for intermediate-age clusters in M83 and two other spiral
galaxies (NGC\,628, NGC\,1313) do not follow the distributions
in M82. This slight difference in behaviour is most likely related to the
lower masses of M82 and LMC/SMC as compared to the other galaxies
with which we compared our data.

Majority of SSCs in M82 follows a mass-radius relation with a 
logarithmic slope of $b=0.29$$\pm$0.05. 
We identify a group of four massive-compact SSCs that are outliers to 
this relation.
We used the semi-analytical cluster evolutionary code, EMACSS, to
understand the observed behaviour of SSCs in the mass-radius space.
We considered a set of more than 80 simulated clusters that cover 
the range of radius, mass and mean stellar densities observed in M82. 
For each of these simulated clusters, we predicted their evolutionary 
trajectory in the mass-radius space. 
From these simulations, we conclude 
that 23\% of the clusters are tidally-limited, with the rest undergoing
expansion at present. Thus, the majority of M82 disk clusters are not tidally
limited, in spite of they following a mass-radius relation with a
logarithmic slope of $b=0.29$, a value close to the value expected for tidally-limited clusters.
Simulations show that the mass-radius relation for a population of expanding clusters,
flattens from the Virial slope of 0.5 at birth. 
The group of massive-compact SSCs
is evolving unaffected by the tidal field of M82, having only a mild (30--50\%) expansion
during the whole evolution. 

Evolving the clusters forward, we find that the majority of clusters tidally truncated 
as well as those currently experiencing large expansion
will dissolve in $\sim$2~Gyr. On the other hand, the group of four massive-compact
SSCs will survive for Hubble time. The end parameters of these
SSCs agree well with the parameters of Galactic GCs, after allowing
for the late-time contraction of the cluster due to core-collapse, which has
not been properly modelled in EMACSS.  
The currently observed central surface brightness and the core radius of these 
four SSCs fall on the fundamental-plane defined by the Galactic GCs, 
after taking into account dimming introduced due to passive evolution
from 100~Myr to 12~Gyr.
These comparisons suggest that 
the group of massive-compact SSCs 
are candidates to proto-GCs. 

We also identified a group of five compact outer-disk SSCs that are not
yet tidally truncated and would survive for the Hubble time. 
These clusters in general loose significant amount of mass during their 
long-term evolution and end up with larger radii and less mass as compared to the
Galactic GCs. Their end values are in excellent agreement with the values
observed for faint fuzzies in galaxies.

\section*{Acknowledgments}
\addcontentsline{toc}{section}{Acknowledgements}

BCO thanks CONACyT for granting PhD research fellowship 
and for the support through the program of
research assistants (grant CB-2019-18276) that enabled
her to carry out the work presented here. We also thank CONACyT for
the research grants CB-A1-S-25070 (YDM), CB-2014-240426 (IP), and
CB-A1-S-22784 (DRG), that allowed the acquisition of a cluster that
was used for computations in this work.  We thank the anonymous referee 
for the valuable comments that improved this work significantly.

\section*{Data Availability}

The data underlying this article are available in the article and in its 
online supplementary material.

\bsp	
\label{lastpage}
\bibliographystyle{mnras}
\bibliography{bibliografia}

\appendix

\section{Velocity dispersion profile for Moffat-EFF profile}\label{App:vel_disp}

\begin{figure}
\begin{center}
\includegraphics[width=\columnwidth]{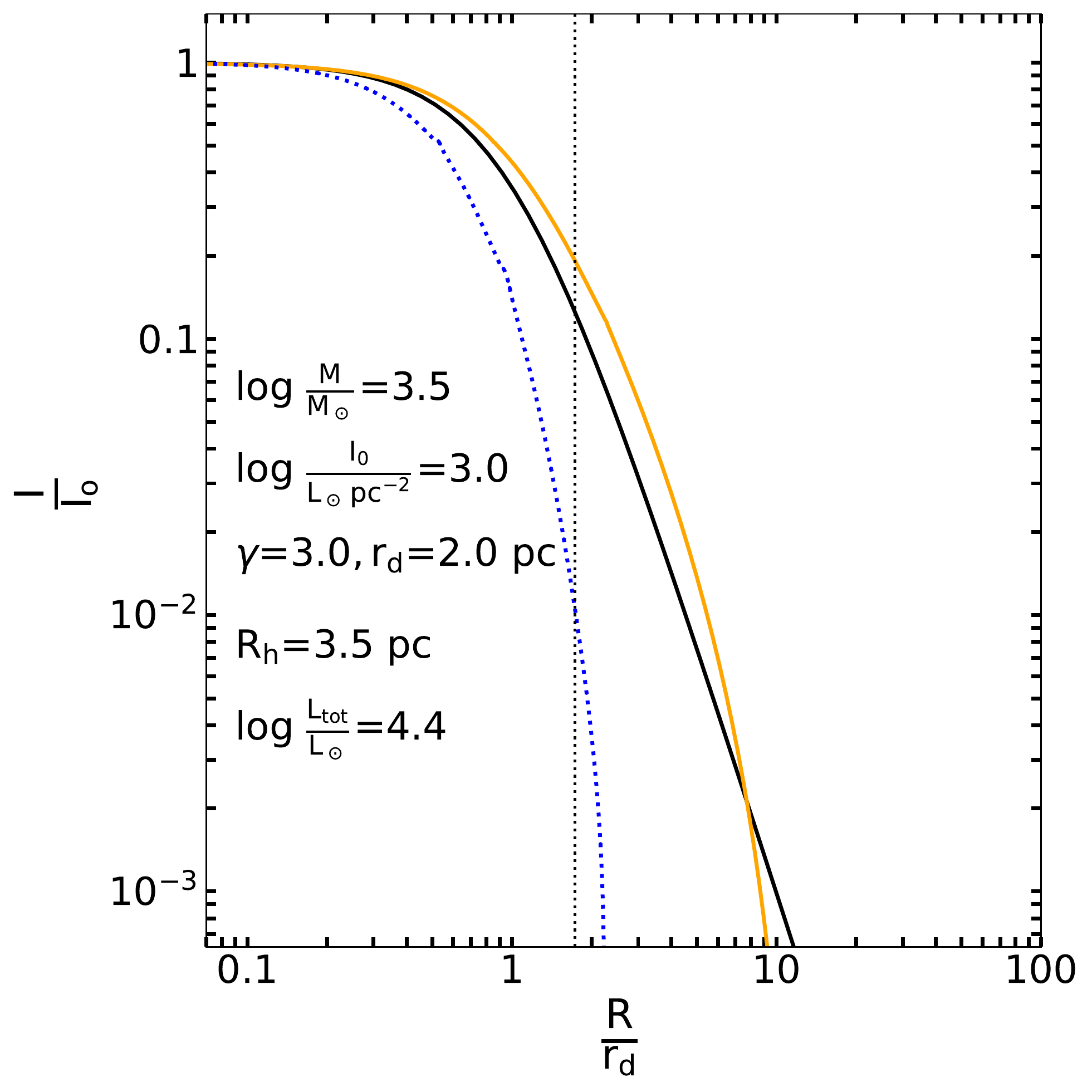}
\includegraphics[width=\columnwidth]{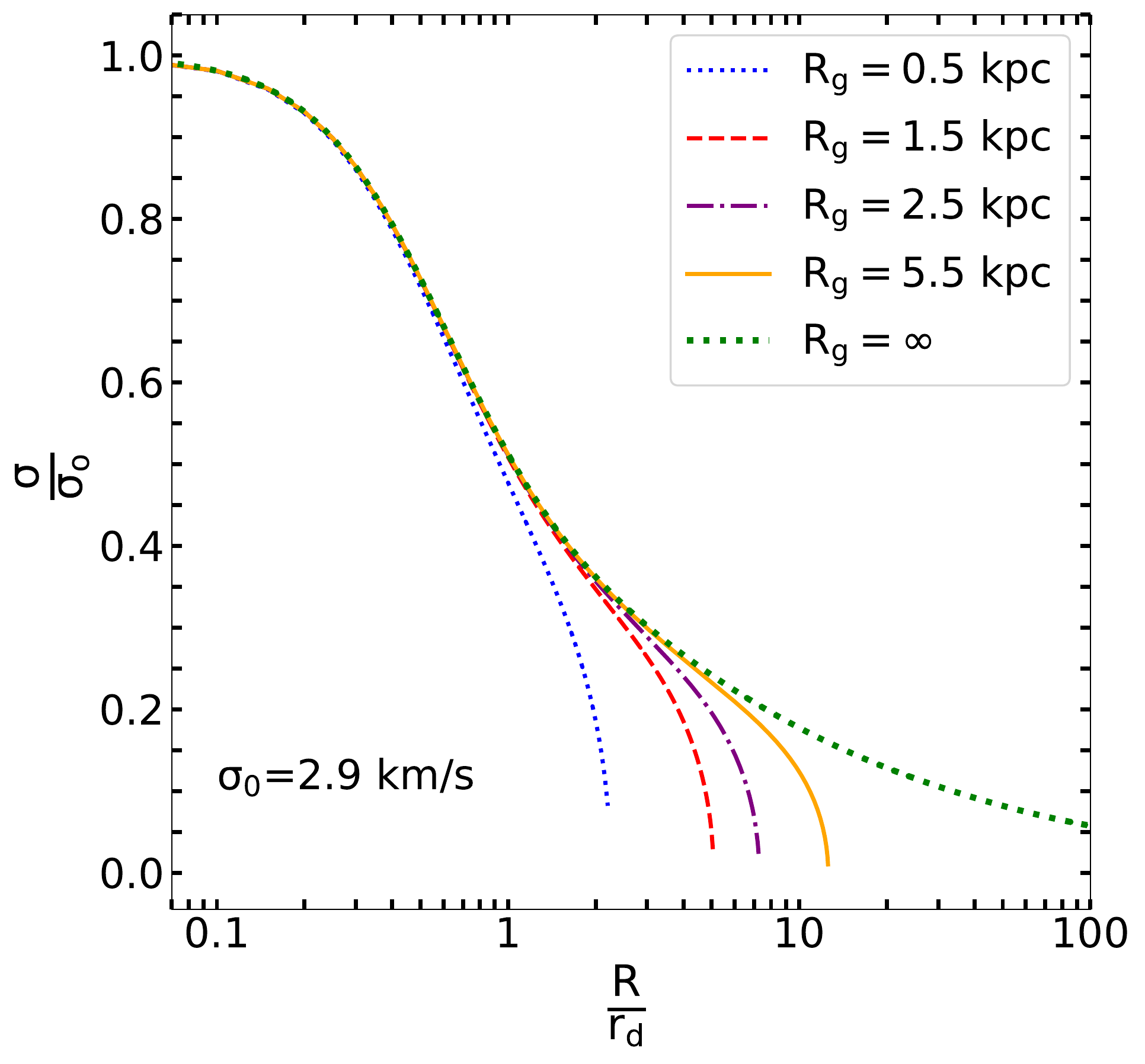}
\caption{(Top) Radial intensity distribution of an illustrative
cluster (black) obeying a Moffat-EFF profile with the properties indicated on
this plot, compared with that of a cluster obeying a King model with the same
core properties, but with the tidal radii the cluster would have if placed
at $R_{\rm g}$=0.5~kpc (blue
line) and 5.5~kpc (orange line) in the disk of M82.
The $R_{\rm h}$ corresponding to Moffat-EFF profile is shown by the
vertical dotted line.
(Bottom) Velocity dispersion profiles for the Moffat-EFF models under the
effect of tidal forces at distinct $R_{\rm g}$ values in the disk of
M82, shown
using the line types indicated in the inset.
Profiles in both the panels are normalised to the corresponding central values.
The mass in bound stars (inside the tidal radius) for this
illustrative case corresponds to 60\% and 92\% of the total mass of
the cluster (integral over the Moffat-EFF profile)
at $R_{\rm g}$=0.5~kpc and 5.5~kpc, respectively.
}
\label{fig:disp_prof}
\end{center}
\end{figure}

We used the solution obtained by \citet{Elson} for a spherical cluster with isotropic velocity distribution
in hydrostatic equilibrium (their equation~16) to calculate the velocity dispersion profile $\sigma(r)$
for a cluster obeying a power-law density profile such as that for the Moffat-EFF profile.
We used a flat rotation curve of v=100~km/s \citep{Greco2012} to present the tidal field 
$\rm 4\Omega^2-\kappa^2=\frac{v^2}{R_{\rm g}^2}$ of M82 at the galacto-centric radius $\rm R_{\rm g}$.
Once $\sigma$(r) is obtained, and thus $\sigma_0$, we proceed to project it into the plane of the sky following the prescription by \citet{Galdynbook}
\begin{equation}
I(R)\sigma^2_p(R)=2\int_R^\infty \frac{j(r)\sigma(r)^2rdr}{\sqrt{r^2-R^2}}
\end{equation}
with $I(R)$ the intensity profile, in terms of the semi-major axis $R$, and $j(r)$ the three-dimensional luminosity density profile.
Moffat-EFF profiles, being power-law in form do not have a cut-off, hence they do not have an implicit tidal radius.  
However, given the tidal field at the location of the cluster, it is possible to define such a radius as the one where the dispersion 
velocity reaches zero, without breaking the hydrostatic equilibrium condition \citep{Elson}.
From the obtained $R_{\rm t}$ values it is possible to compute the bound mass of the clusters $\rm M_{bound}$, by 
integrating $\Sigma(R)$ in the limits between 0 and $R_{\rm t}$. This integration has an analytical solution given by
\citep{Elson},
\begin{equation}\label{eqn:Mbound}
\frac{M_{\rm bound}}{M} = 1-\left[1+\left(\frac{R_{\rm t}}{r_{\rm d}}\right)^2\right]^{1-\gamma/2},
\end{equation}
where $M=\Gamma\times L_{\rm tot}$ is the total mass and $\Gamma$ is the mass-to-light ratio. The $R_{\rm t}$, 
$\frac{M_{\rm bound}}{M}$, $M$ and $\sigma_{\rm p,0}$ are given in columns~6, 7, 10 and 14 of Table~\ref{tab:der_pars}, respectively.

In Fig.~\ref{fig:disp_prof}, we show the distribution of the intensity and $\sigma$(r) profiles for an 
illustrative cluster resembling the properties of the sample M82 disk SSCs.  The cluster has a low mass
and a radius as well as a $\gamma$ values close to the median radius and Moffat-EFF index values 
of the sample (3.2$\times10^3$~M$\odot$, $r_{\rm d}$=2.0~pc, $\gamma$=3.0). 
The y-and x-axis of the plots are shown normalized to the central
values, and $r_{\rm d}$, respectively. The $\sigma$(0) value for the cluster is 2.9~km/s.
The $\sigma$(r) profiles are shown under the influence of tidal fields
at various $R_{\rm g}$ values. The profile corresponding to $R_{\rm g}=\infty$ corresponds to an isolated case, where
the velocity never reaches zero for a final $R_{\rm g}$.
It can be seen that the velocity abruptly falls to zero in the presence of a tidal field, with the radius at which
it reaches zero (tidal radius), progressively smaller at smaller values of $R_{\rm g}$.
The $\sigma$(r) profiles for the cluster remains almost
constant for $r<r_{\rm d}$, implying the cluster core is unaffected by the tidal fields \citep{Elson}.

\section{Simulations using EMACSS}\label{App:EMACSS}

\begin{figure}
\begin{center}
\includegraphics[width= \columnwidth]{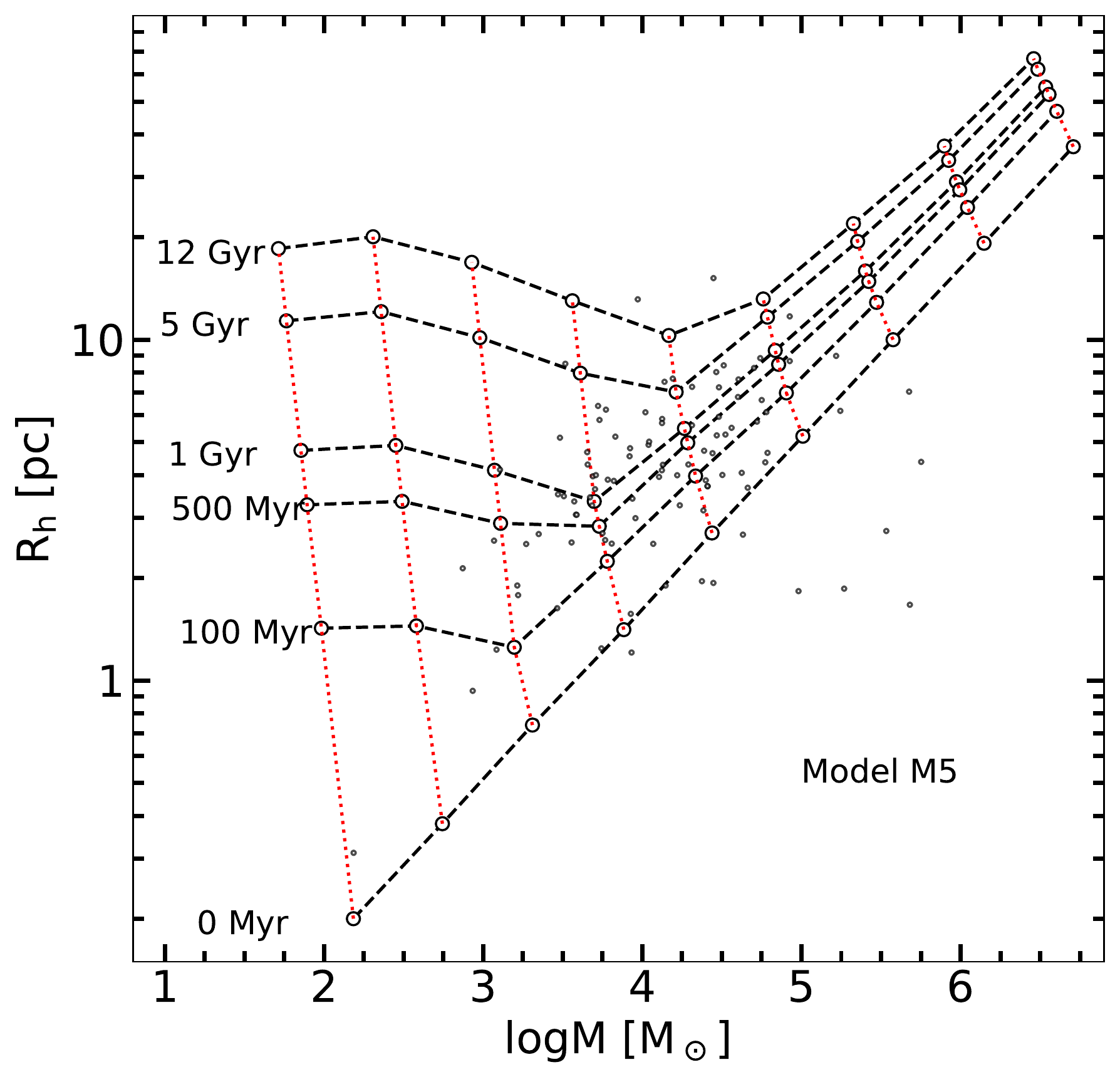}
\caption{Evolution of the mass-radius relationship over 0, 100, 500 Myr 1, 5 and 12 Gyr, using the fast evolution code Evolve Me a Cluster of Stars \citep[EMACSS][]{AlexanderGieles2012,AlexanderGieles2014} for an isolated cluster (dashed lines) from an initial virial mass-radius relation (Model M5 in Tab. \ref{tab:emacss}).  The vertical red dotted lines join every point from its initial condition to its later stage.  The observed values of the M82 disk SSCs are shown with black points.}
\label{fig:ap_sim_emacss}
\end{center}
\end{figure}

\begin{figure*}
\begin{center}
\subfloat{\includegraphics[width= 0.5\textwidth]{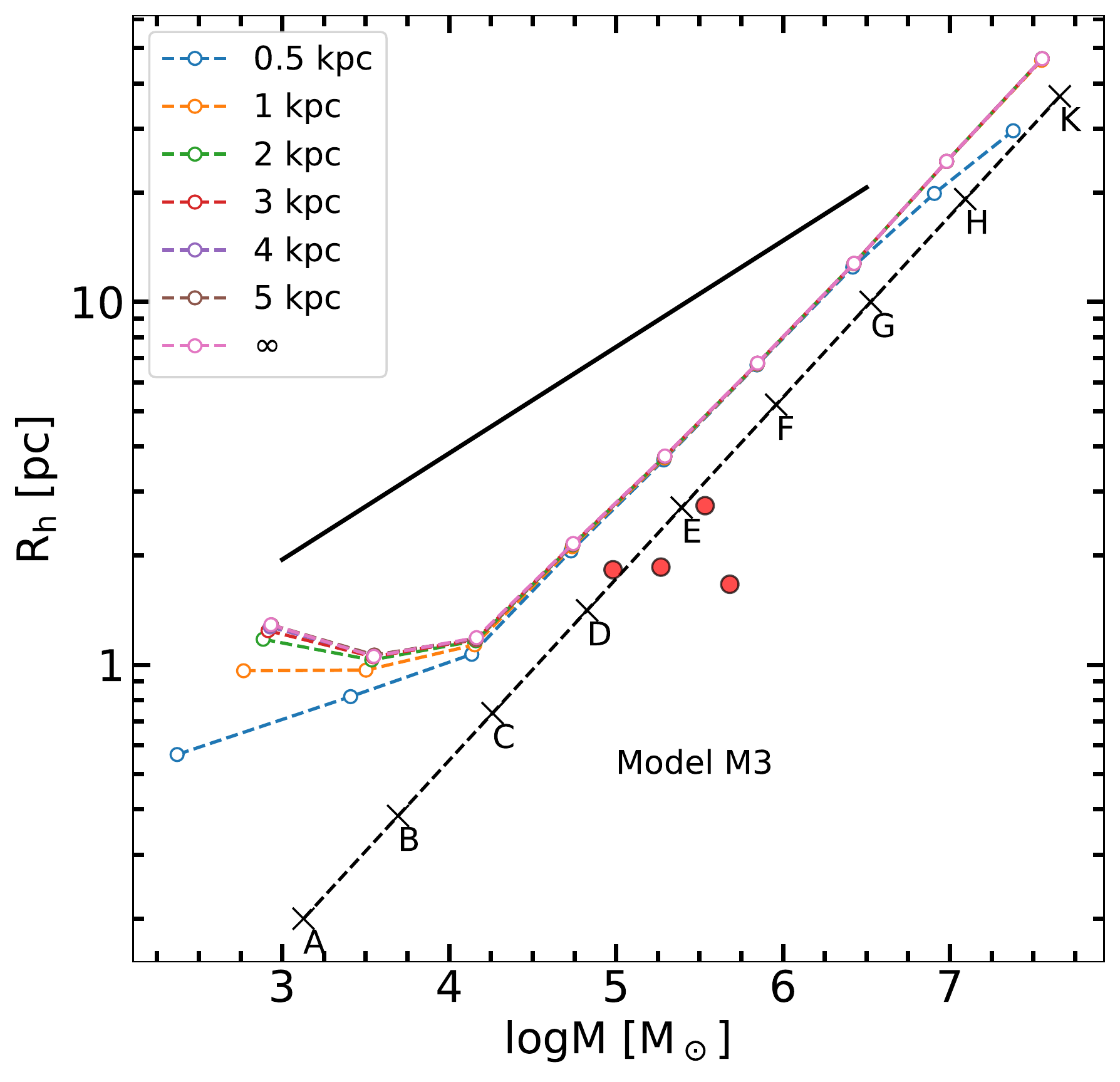}}
\subfloat{\includegraphics[width= 0.5\textwidth]{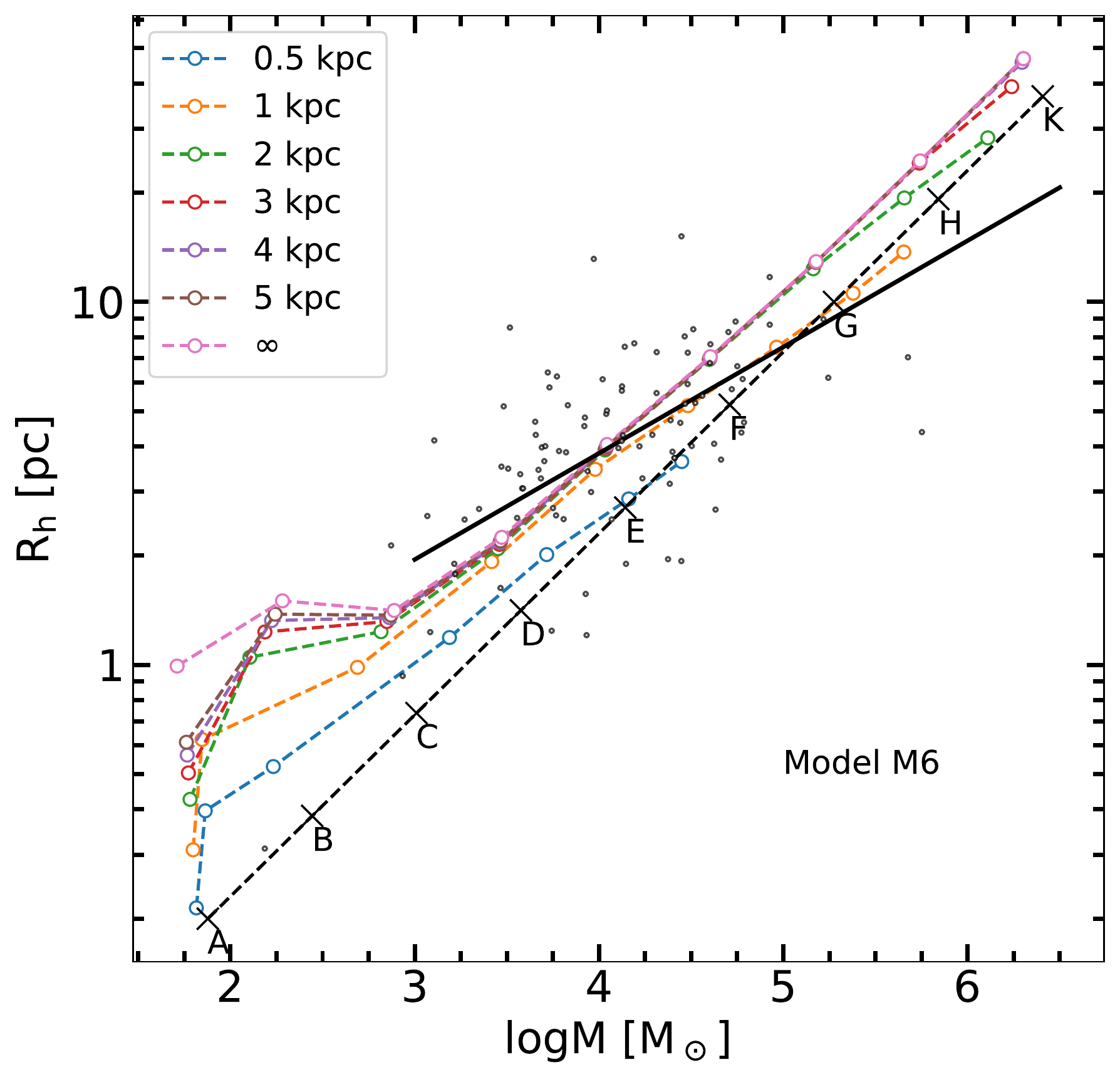}}
\caption{Evolution of the mass-radius relationship over the first
100~Myr using the fast evolution code EMACSS 
\citep{AlexanderGieles2012,AlexanderGieles2014} for clusters represented 
by a singular isothermal halo of constant mean stellar density,
at different galactocentric radii $R_{\rm g}$ shown in the inset box.
The value of $\infty$ corresponds to isolated clusters.
The initial mass-radius relation is shown by the dashed lines,
with letters A--K corresponding to the models described in Tab.~\ref{tab:emacss}.
The best-fit line for majority of M82 SSCs (shown in small black circles) is shown by the
solid line.
In the left panel, we show the evolution for model M3 (high density)
and on the right panel that for model M6 (density similar to that of the majority of M82 SSCs). 
Clusters that are represented by M3 C to F evolve in an identical way
at all galactocentric radii. These clusters are in their expansion phase,
and have not yet reached their tidal radius. On the other hand, highly extended
($R_{\rm h}>$5~pc) and low-mass ($M<10^4$~M$_\odot$) clusters start getting
truncated at $R_{\rm g}$=0.5~kpc. None of these models reach the zone where
majority of the observed points is located. On the other hand, several
cases in the M6 model are tidally-truncated for $R_{\rm g}<$2.0~kpc.
Distribution of these truncated models closely matches the observed 
distribution of points.  The group of massive-compact SSCs is indicated in 
red circles in the upper panel.  The rest of the observed points is indicated with
black points in the bottom panel.}
\label{fig:ap_sim_emacss2}
\end{center}
\end{figure*}

We have performed a set of simulations in order to understand the possible 
evolution of the mass-radius relation with the conditions of M82.  We have 
simulated the evolution of clusters in isolation, following an initial 
virial mass-radius relation \citep{Gieles2010} and also under the
gravitational potential of M82, which is represented by a flat rotation 
curve with a circular velocity of 100~km s$^{-1}$ \citep{Greco2012} corresponding
to a singular isothermal halo.
The evolutionary results are saved for the following ages: 0 Myr, 100 Myr, 500 Myr, 
1 Gyr, 5 Gyr and 12 Gyr.
We have considered 6 different values of galacto-centric radii ($R_{\rm g}$) 
from 0.5 to 5 kpc, intended to cover the M82 disk SSCs $R_{\rm g}$ values.  
In the main text, we show only the results for isolated clusters
and those at $R_{\rm g}$=2~kpc at 100~Myr, for preserving
the clarity of the figure.

We present a few plots that allow us to illustrate
the evolutionary behaviour at different times (Fig. \ref{fig:ap_sim_emacss}) and 
at different $R_{\rm g}$ values (Fig. \ref{fig:ap_sim_emacss2}) .

\end{document}